\documentclass[fleqn,usenatbib]{mnras}

\usepackage{newtxtext,newtxmath}
\usepackage{subfig}
\usepackage[T1]{fontenc}

\DeclareRobustCommand{\VAN}[3]{#2}
\let\VANthebibliography\thebibliography
\def\thebibliography{\DeclareRobustCommand{\VAN}[3]{##3}\VANthebibliography}

\usepackage{graphicx}	% Including figure files
\usepackage{amsmath}	% Advanced maths commands
\title[Library of Semi-Empirical MILES Stellar Spectra with Variable Abundances]{sMILES: A Library of Semi-Empirical MILES Stellar Spectra with Variable [$\alpha$/Fe] Abundances}
\author[A. T. Knowles et al.]{
Adam T. Knowles$^{1,2}$\thanks{E-mail: adamtknowles@gmail.com}, A. E. Sansom$^{1}$, C. Allende Prieto$^{3,4}$, A.Vazdekis$^{3,4}$
\\
$^{1}$Jeremiah Horrocks Institute, School of Physical Sciences and Computing, University of Central Lancashire, Preston, PR1 2HE, UK\\
$^{2}$ Departament de F\'isica, Universitat Polit\'ecnica de Catalunya, c/Esteve Terrades 5, 08860 Castelldefels, Spain\\
$^{3}$Instituto de Astrof\'isica de Canarias, V\'ia L\'actea, 38205 La Laguna, Tenerife, Spain\\
$^{4}$Universidad de La Laguna, Departamento de Astrof\'isica, 38206 La Laguna, Tenerife, Spain\\
}

\date{Accepted XXX. Received YYY; in original form ZZZ}

\pubyear{2020}

\begin{document}
\label{firstpage}
\pagerange{\pageref{firstpage}--\pageref{lastpage}}
\maketitle

\begin{abstract}
We present a new library of semi-empirical stellar spectra that is based on the empirical MILES library. A new, high resolution library of theoretical stellar spectra is generated that is specifically designed for use in stellar population studies. We test these models across their full wavelength range against other model libraries and find reasonable agreement in their predictions of spectral changes due to atmospheric $\alpha$-element variations, known as differential corrections. We also test the models against the MILES and MaStar libraries of empirical stellar spectra and also find reasonable agreements, as expected from previous work. We then use the abundance pattern predictions of the new theoretical stellar spectra to differentially correct MILES spectra to create semi-empirical MILES (sMILES) star spectra with abundance patterns that differ from those present in the Milky Way. 
The final result is 5 families of 801 sMILES stars with [$\alpha$/Fe] abundances ranging from $-$0.20 to 0.60 dex at MILES resolution (FWHM=$2.5\,${\AA}) and wavelength coverage ($3540.5-7409.6\,${\AA}). We make the sMILES library publicly available.
\end{abstract}


%%%%%%%%%%%%%%%%%%%%%%%%%%%%%%%%%%%%%%%%%%%%%%%%%%

%%%%%%%%%%%%%%%%% BODY OF PAPER %%%%%%%%%%%%%%%%%%
\section{Introduction}
The stellar populations within a galaxy hold information about how that system formed and evolved. Contained within its integrated light, the abundances of various chemical elements in the atmospheres of its constituent stars provide insights into a galaxy's past. Comparing observations of galaxies to models can reveal these clues and allows for the determination of stellar population properties including age, metallicity, chemical abundances and star formation history, all of which provide details about their formation and evolution. The framework for such models was developed by Tinsley (\citealt{Tinsley68}; \citealt{Tinsley80}), in which the time-evolution of stellar population colours and chemical abundances were predicted and matched to observations. These first models provided the basis of modern evolutionary stellar population synthesis (SPS) that is widely-used to fit spectral indices or full spectra of unresolved populations in external galaxies (e.g. \citealt{Bruzual83}; \citealt{Worthey94}; \citealt{Vaz96,Vaz99Model,Vaz2010,Vaz2015}; \citealt{Coelho07}; \citealt{Conroy2012a}).

A key component in the generation of SPS models is the stellar library used to convert the predictions of stellar evolutionary calculations, values of surface gravity (log g) and effective temperature ($\textrm{T}_{\textrm{eff}}$) at different metallicities, into spectra. An effective library would contain stars of various evolutionary stages, covering a large range of \textrm{$\textrm{T}_{\textrm{eff}}$, log g} and metallicity (e.g. characterised by [Fe/H]\footnote{[A/B]=$\log[{n(A)/n(B)}]_{*}$ - $\log[{n(A)/n(B)}]_{\odot}$, where $n(X)/n(Y)$ is the number abundance ratio of element A, relative to element B.}). More recent work has also covered abundance patterns (e.g. [Mg/Fe] in \citealt{Milone2011} and [$\alpha$/Fe] in \citealt{Yan2019}).

 Stellar libraries can consist of theoretical spectra (e.g. \citealt{Coelho2014}) or observed spectra (e.g. \citealt{SanchezBlazquez2006}). Theoretical spectra have the advantage of covering a wide parameter space and do not have the typical observational limitations. They are however limited by the calculations, which embrace multiple simplifying physical assumptions. The treatment of convection, microturbulence, atmospheric geometry, local thermodynamic equilibrium (LTE) are all choices that limit their accuracy. Several theoretical libraries that cover a wide parameter range have been produced for large spectroscopic surveys or Single Stellar Population (SSP) modelling (\citealt{Coelho05}; \citealt{Coelho2014}; \citealt{Bohlin17}; \citealt{Allende18}). SPS models computed using only theoretical spectra have been used in the literature (e.g. \citealt{Maraston05}; \citealt{Coelho07} at low and high resolution, respectively). 

Although observational spectra correctly represent all of the physics and spectral features present in stars, they suffer from observational constraints such as limited wavelength coverage, spectral resolution, atmospheric absorption, emission residuals from the sky and noise. A major issue affecting empirical libraries is the limited parameter space covered, which is unavoidable because spectra are drawn from samples of stars in the vicinity of the solar neighbourhood and will therefore be representative of the Milky Way's chemical evolution. It is possible to obtain spectra of stars from further distances with differing chemical abundance patterns, but long exposure times limit these observations to only small, bright samples. A historical review of empirical libraries is presented in \cite{Trager2012}. A very popular empirical library is the Medium-resolution Isaac Newton Telescope Library of Empirical Spectra (MILES) (\citealt{SanchezBlazquez2006}) which consists of $\sim$1000 flux calibrated stars between $3500-7500\,${\AA}. Coverage of empirical libraries in effective temperature and surface gravity is good. However, the abundance patterns sampled in the solar neighbourhood are limited, constraining the range of stellar populations that can be accurately modelled. The abundance patterns of other galaxies and even within our Galaxy are not always the same as the solar neighbourhood (e.g. \citealt{Edvardsson93}; \citealt{Holtzman2015}). These empirical libraries, and the SSP models that can be generated from them, are therefore limited. Examples of SSP models computed using empirical stars can be found in \cite{Vaz99Model} and \cite{Vaz2010}. 

Another approach is to use combinations of empirical and theoretical spectra to increase the wavelength coverage of stellar population models (e.g. \citealt{Bruzual03}; \citealt{Maraston11}). An analysis of the impact of using theoretical or empirical stellar spectra in the generation of stellar population models is presented in \cite{Coelho20}. 

The elemental abundance patterns of galaxies highlight the time-scales in which their constituent stellar populations were formed. Even moderate resolution spectra contain details that allow for measurements of individual chemical abundances (e.g. R$\sim$2000 in \citealt{Parikh19}, using MaNGA \citealt{Blanton17}). A useful abundance ratio to measure is [$\alpha$/Fe], because the sources and time-scales of interstellar medium (ISM) enrichment for $\alpha$-capture and iron-peak elements are different. The ISM is polluted with $\alpha$-elements by Type II supernovae on shorter time-scales than iron-peak elements that mostly originate from Type Ia supernovae. The overabundance of [Mg/Fe] compared to the solar neighbourhood observed in early-type galaxies (ETGs) is usually attributed to short formation time-scales (e.g. see the review of \citealt{Trager98} and references therein).

If one can quantify how stellar spectra are sensitive to elemental abundances it is possible to build stellar spectral libraries, and therefore SPS models, which contain abundance patterns different from the solar neighborhood. This is motivated by the different abundance patterns seen in external systems such as ETGs and Dwarf Spheroidal galaxies (dSphs) (e.g. see the review of \citealt{Conroy13} and references therein, in addition to \citealt{Letarte2010}; \citealt{Conroy2014}; \citealt{WortheyTangServen2014}; \citealt{Sen2018}).

 To account for non-solar abundance patterns in SSP models, a hybrid approach can be made in which a combination of the predictions from theoretical spectra (calculated from stellar spectral models or fully theoretical SSP models) and the accuracy of empirical spectra is used. A prediction of how abundances affect spectral lines is applied to empirical spectra to account for different abundance patterns, known as a differential correction. These corrections can be performed on specific spectral lines, presented in the form of response functions (e.g. \citealt{Tripicco1995}; \citealt{Korn2005}), or can be calculated for the full spectrum. SSP models can be generated using a fully empirical library as the base, with differential corrections made from theoretical models to account for different abundance patterns. Some of the first works to take a differential abundance pattern approach in full spectrum SPS modelling were that of \cite{Prugniel07} and \cite{Cervantes07}, followed by \cite{Walcher09}. This work was then expanded by \cite{Conroy2012a}, who calculated the response of SSP spectra for element abundance variations, at fixed metallicity, near the solar value.

This method of using the abundance pattern predictions of models can be applied to individual stars in empirical libraries, which are then used to generate SSP spectra (e.g. \citealt{LaBarbera17} for [Na/Fe] variations), or to fully empirical SSP spectra directly from fully theoretical SSP models (e.g. \citealt{Vaz2015} for [$\alpha$/Fe] variations). In this work we build a stellar spectral library with stars that contain atmospheric abundances that can encompass a range of extragalactic environments. We use state-of-the-art theoretical spectra and apply their abundance predictions to existing empirical MILES stars. The result is a library of semi-empirical star spectra, covering a broad range of stellar parameter space, including [$\alpha$/Fe] variations spanning a larger range and finer sampling than previously computed \cite{Vaz2015} SSP models. Our aim is to produce a database of stars with different abundance patterns, which can then be directly used in the construction of new SSP models. We make the semi-empirical stellar library available for public use in both population synthesis and stellar applications. We chose to base the semi-empirical library on the widely-used MILES empirical library for which SSP modelling methods already exist.

The structure for this paper is as follows. Section~\ref{sec:ModelSpectra} describes the generation and processing of a new theoretical stellar library, for use in stellar population modelling. Section~\ref{sec:ModelTesting} tests this new library through comparisons to other published theoretical libraries. Section~\ref{sec:MILESstars} outlines the underlying empirical MILES stellar library used in the calculations and their parameters that we adopt. Section~\ref{sec:sMILESstars} describes the interpolation to create theoretical MILES stars, plus the differential correction process used in the creation of semi-empirical MILES star spectra with different [$\alpha$/Fe] abundances. Section~\ref{sec:TestingObs} tests the star spectra through comparisons to observations. Section~\ref{sec:Summary} presents our summary and conclusions.

\section{Models of Stellar Spectra}
\label{sec:ModelSpectra}
To address the limitations of using purely empirical stellar spectra in SSP models, we use theoretical spectra with varying abundance patterns. By taking ratios of theoretical spectra and applying them to existing MILES stars, we create a library of semi-empirical MILES star spectra with different [$\alpha$/Fe] abundances that can be used to compute semi-empirical SSPs. This approach, making use of both models and observations, builds upon the work of \cite{LaBarbera17}, implementing both the accuracy of empirical spectra with the differential abundance pattern predictions of theoretical spectra. Using only differential predictions from theoretical spectra has been shown to reproduce observations of abundance pattern effects more accurately than fully theoretical spectra, particularly for wavelengths below $\textrm{Mg}_{\textrm{b}}$ (e.g see figure 11 of \citealt{Knowles19} or \citealt{Martins07}; \citealt{Bertone08}; \citealt{Coelho2014}; \citealt{Villaume17}; \citealt{Allende18}).

This approach requires a theoretical library of stellar spectra, from which abundance pattern predictions are used. Rather than use an existing  library that has particular stellar parameter and abundance pattern coverage as well as a wavelength range for use in a specific application, such as the H band investigated using APOGEE (e.g. \citealt{Zamora2015}), we compute a new, high resolution theoretical stellar spectral library that is specifically designed for this project. 

In \cite{Knowles19} we tested theoretical spectra from three different groups of modellers, who used different software, and found that the differences between models is less than the differences between models and observations. Therefore, for this work we chose a method with which we were most familiar and that achieved good results in the comparisons to observations. Based on the results obtained from testing in \cite{Knowles19}, we follow the calculation method presented in detail in \cite{Mezaros2012} and \cite{Allende18}. This section summarises the computation methods and parameter choices for our new library, covering UV to near infrared wavelengths.

\subsection{Computation Method}
\label{sec:CompMethod}
The production of theoretical stellar spectra requires two main processes: model atmosphere calculation, followed by radiative transfer through an atmosphere to produce an emergent spectrum, requiring the use of a synthetic spectrum code together with appropriate opacities, including a list of atomic and molecular absorption transitions and a specification of element abundances. The self-consistent approach would be to exactly match the chemical abundances in both stages of the production. To reduce computation time, a simplification is typically made in which only the dominant sources of opacity are varied in the model atmosphere whilst more elements are varied in the detailed synthetic spectrum calculation. 

The model atmospheres used in this project were generated using ATLAS9 (\citealt{Kurucz1993}), for which recently computed opacity distribution functions (ODFs) already existed. These ODFs cover the main sources of line opacity variations in stellar atmospheres, including variations in metallicity, $\alpha$-element and carbon abundances. The ODFs and model atmospheres used in this paper are described in \citealt{Mezaros2012}. They are publicly available\footnote{\url{http://research.iac.es/proyecto/ATLAS-APOGEE//}} and were used in the APOGEE analysis pipeline (\citealt{Ana2016}). The $\alpha$ elements we included are: O, Ne, Mg, Si, S, Ca and Ti. The ODFs and model atmospheres used in this work adopt \citealt{Asplund2005} solar abundances and a microturbulent velocity of $2\,\mathrm{km\,s^{-1}}$. We note here that this fixed value of microturbulent velocity is only used in the atmospheric model generation. In the spectral synthesis stage we use a microturbulent velocity that is dependent on effective temperature and surface gravity. The model atmospheres consist of 72 plane parallel layers from $\log\tau_{\mathrm{Ross}}=-6.875$ to $+0.200$ in steps of $\Delta\log\tau_{\mathrm{Ross}}=0.125$ (\citealt{Castelli03}). Alternative model atmosphere calculation methods would include the opacity sampling regimes of both MARCS (\citealt{Gustafsson75}; \citealt{Plez92}; \citealt{Gustafsson08}) and ATLAS12 (\citealt{Kurucz2005a}; \citealt{Castelli2005a}) and have been found to produce similar predictions to ATLAS9 models (e.g. \citealt{bon11}; \citealt{Mezaros2012}; \citealt{Knowles19} ).

The stellar atmospheres used in this work have metallicities ranging from $-$2.5 to 0.5, for a range of carbon and $\alpha$ abundances presented later in this section, covering a large section of the MILES empirical stellar library. This region is deemed reliable to interpolate within when computing stellar population models, given the distribution of MILES stars (see figure 10 of \citealt{Milone2011}). This reliability is expressed through a Quality Number (Q$_n$), defined in \citealt{Vaz2010}) and shown in figure 6 of \cite{Vaz2015}. Q$_n$ gives a quantifiable measure of SSP spectra reliability, based on the density of stars around isochrone locations used in SSP calculations, with higher densities resulting in larger Q$_n$ values.

For the radiative transfer stage of this work, we use ASS$\epsilon$T (Advanced Spectrum SynthEsis Tool) (\citealt{Koesterke2009}). ASS$\epsilon$T is a package, consisting of Fortran programs, providing fast and accurate calculations of LTE and non-LTE spectra from 1{\small D} or 3{\small D} models. Ideally we would calculate cool star models with 3D geometry and account for NLTE. However, we note that 1{\small D}, LTE modelling normally handles the opacity in more detail than in existing 3{\small D} and NLTE codes, which are computationally costly (e.g. \citealt{bon11}). Therefore, we caution that our 1D, LTE models will be increasingly poorer representations of real stars at lower temperatures, below about 4000 K. Future work might investigate whether more complex models would produce better estimates of differential element responses in the spectra of the coolest stars.
 For generating a large number of theoretical spectra, each covering a broad wavelength range, we use the 1{\small D} and LTE mode of ASS$\epsilon$T, with the input ATLAS9 atmospheres, to produce a library of synthetic spectra at air wavelengths. Calculations were done in “ONE-MOD” mode within ASS$\epsilon$T, with the opacities computed exactly for each model at every atmospheric depth. Several important aspects of the models are summarised below.
\begin{itemize}
\item{\textbf{Solar Abundances} -  To maintain abundance consistency in the computation, we define abundances relative to \cite{Asplund2005} solar abundances in both ATLAS9 and ASS$\epsilon$T.}
\item {\textbf{Abundance Definitions} - The models were computed with variable metallicity ([M/H]), ([$\alpha$/M]) and carbon ([C/M]) abundances. [M/H] here is defined as:
\begin{equation}
\textrm{[M/H]}=\log[{n(M)/n(H)}]_{*} - \log[{n(M)/n(H)}]_{\odot}, 
\label{MHDefEq}
\end{equation}
where $n(M)$ is the number of nuclei of any particular element with atomic number greater than two, but not the summation of all, i.e. it applies to iron, lithium, potassium, and any single element. [M/H] here is therefore defined as a scaled-metallicity in which all metals, apart from the $\alpha$-elements and carbon if they are also non-solar, are scaled by the same factor from the solar mixture (e.g. [M/H]=0.2=[Fe/H]=[Li/H]). This definition means [$\alpha$/M]=[$\alpha$/Fe] and [C/M]=[C/Fe].}
\item {\textbf{ODFs} - To avoid complex computation of new ODFs with variable abundances, we generate models on a grid for which ODFs existed. Therefore, we are constrained to generate synthetic spectra on the existing grid points from \cite{Mezaros2012}. These grid points dictate the abundance pattern sampling of the current library.}
\item{\textbf{Line lists} - The line lists used in the calculations are described in detail in \cite{Allende18}. In summary, metal and molecular transitions are obtained from Kurucz\footnote{\url{http://kurucz.harvard.edu/}}. Molecules present in the calculation include H$_2$, CH, $\textrm{C}_{2}$, CN, CO, NH, OH, MgH, SiH, and SiO. TiO transitions are only included for stars below 6000{\small K}, as explained in Section~\ref{sec:NewGrids}}.
\end{itemize}

Models were computed at the grid points described in Section~\ref{sec:NewGrids}. The wavelength range of the models was guided by the starting value of the extended MILES library ($\sim1680\,${\AA}) (\citealt{Vazdekis16}) and the inclusion of calcium triplet (CaT) features (at 8498, 8542 and $8662\,${\AA}), to allow for investigation of IMF variations in ETGs. This results in a high resolution theoretical library that is generated spanning the wavelength range of 1680-$9000\,${\AA}. However, for the semi-empirical library, we will be limited to producing semi-empirical stellar spectra with the current MILES library wavelength range of 3500-$7500\,${\AA} as described in Section~\ref{sec:MILESstars}.
\subsection{Element Abundance Variation}
\label{sec:ElementVariations}
The total number of models generated is based on the number of elements varied, their range of variation and number of steps taken, as well as the sampling in other stellar parameters. We specify what element groups are varied in each component of the model computation.
\begin{itemize}
\item \textbf{Model Atmosphere (ATLAS9)} - [M/H], [$\alpha$/M] and [C/M]
\item \textbf{Radiative Transfer (ASS$\epsilon$T)} -  [X/H], where X can be any element from atomic number 2 to 99
\end{itemize}
 Variation of elements is driven by the ODFs and by observations of abundance patterns in external systems (e.g. see \citealt{WortheyTangServen2014}; \citealt{Sen2018}). Therefore, we vary the abundances in the following way.
\begin{itemize}
\item{\textbf{[M/H]} from $-$2.5 to +0.5 in steps of 0.5 dex - where [M/H] is defined in equation (~\ref{MHDefEq}})
\item {[\boldmath{$\alpha$}/\textbf{M}] from $-$0.25 to +0.75 in steps of 0.25 dex (where $\alpha$ = O, Ne, Mg, Ca Si, S and Ti to be consistent with the model atmosphere variations)}
\item {$[\textbf{\textrm{C}/\textrm{M}}]$ from $-$0.25 to +0.25 in steps of 0.25 dex) - carbon abundance has a large impact on stellar spectra. Its atmospheric composition, relative to oxygen, can lead to carbon stars. The balance of C and O is important in the molecular equilibrium of cool stars and the entire atmospheric structure changes significantly when C/O approaches one, producing carbon stars (\citealt{Mezaros2012}; \citealt{Gonneau2016}). With ODFs computed with carbon variations, it is possible to consistently change carbon in both model atmosphere and spectral synthesis components.}
\end{itemize}
Other elements variations that could be synthesised and would be useful in stellar population studies include nitrogen and sodium. However, in this work we present the first stage of this stellar library and focus on $\alpha$ and carbon variations, which are known to have the largest impact on stellar structure and on stellar spectra when changes in their ratios to iron are considered. Considering these two will lead to significant improvements in fitting the spectra of stars and stellar populations. Sodium variations have been considered in \cite{LaBarbera17} at the star and SSP level for a limited number of models using the same methods described here, albeit with abundance variations made only in the radiative transfer component of computation.
\subsection{Microturbulence}
\label{sec:Microturbulence}

\begin{figure*}
    \centering
    \subfloat{{\includegraphics[width=8.5cm]{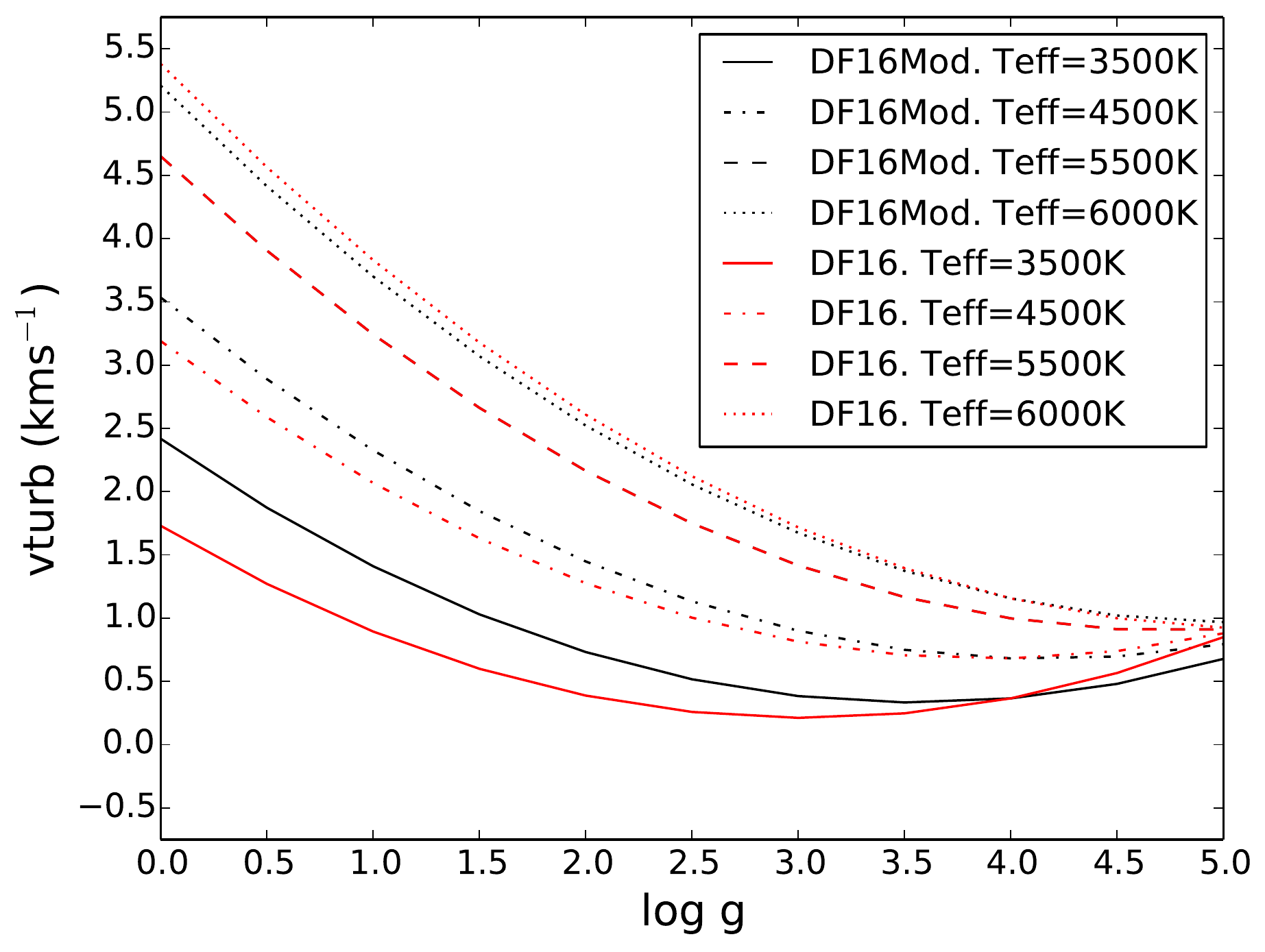}}}
    \qquad
    \subfloat{{\includegraphics[width=8.5cm]{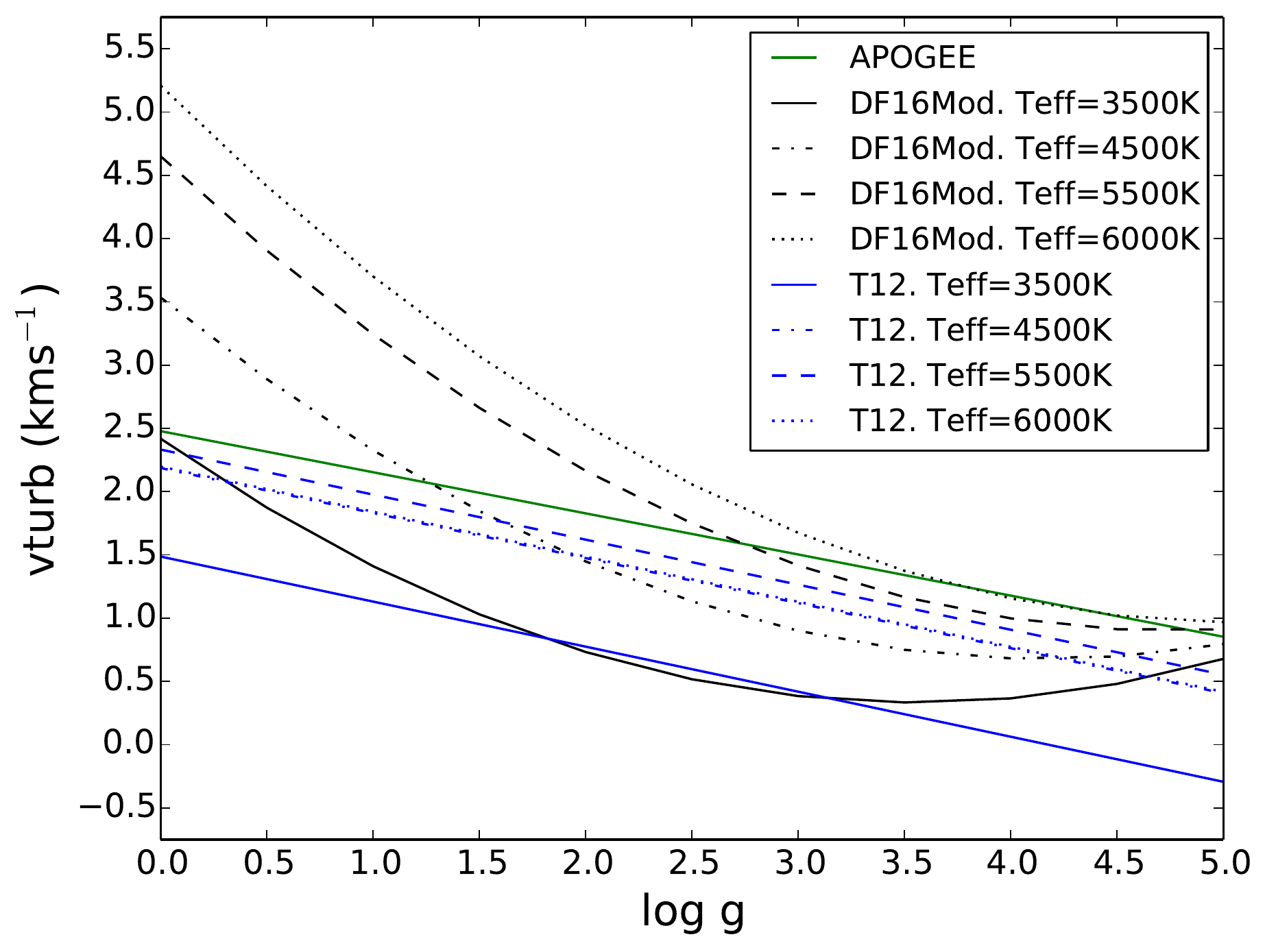}}}
    \caption{Left Panel: Microturbulent velocity as a function of log g for the original \citealt{DF2016} equation (DF16 Red lines) and the modified version of the equation (DF16Mod black lines) for four values of $\textrm{T}_{\textrm{eff}}$. For $\textrm{T}_{\textrm{eff}}$=5500{\small K}, DF16=DF16Mod. The main difference can be seen at lower temperatures, where DF16Mod avoids dropping to such low values of microturbulent velocity. This modification better represents the trends found by observations (e.g. \citealt{Ramirez2013}, their figure 5). Right Panel: Microturbulent velocity as a function of log g for the modified DF16 equation (DF16Mod. Black lines), the \citealt{Thygesen2012} equation (T12. Blue lines) and the APOGEE equation (APOGEE. Green lines). Although the T12 equation appears to follow the linear behaviour of the APOGEE calibration well, problems arise at higher $\textrm{T}_{\textrm{eff}}$  where the equation does not reach the higher values of microturbulent velocity observed at low log g.}
    \label{fig:DF16_old_vs_new_plus_APOGEE_vs_DF16Mod_vs_T12}
\end{figure*}

An important parameter in the computation of one dimensional stellar spectra is the microturbulent velocity. Due to a limitation in classical 1{\small D} models to fully treat the velocity fields present in stellar photospheres correctly, microturbulence is included to match the observed broadening of spectral lines (e.g. \citealt{Struve34}; \citealt{vanParadijs72}). Treated as motions of mass below the mean free path of photons, microturbulence is usually modelled as a Gaussian distribution of velocity dispersion, which in turn produces Doppler shifts that mimic the effect of thermal motions. For weak lines that have typically Gaussian profiles, the effect of microturbulence is to increase the width and reduce the depth of the absorption line, producing no change in equivalent width. However, for stronger and saturated lines for which absorption can occur in the damping wings of line profiles, microturbulence expands the wavelength range of possible absorption and results in reduced saturation and therefore increases the total absorption. Therefore, the choice of this parameter is important because it can affect the resulting line-strengths when calculating synthetic spectra. Although the available ODFs, and therefore model atmospheres, were computed at $2\,km\,s^{-1}$, microturbulent velocity can be varied in ASS$\epsilon$T and therefore we considered the effect of this parameter on the theoretical grid.
The effects of microturbulence on the absolute and differential application of theoretical line-strengths are discussed in \citealt{Knowles19}. The results of these tests are summarised here. 

In general, we found that absolute differences in line-strength indices can be large even for relatively small differences in the adopted microturbulent velocity (of $1\,\mathrm{km\,s^{-1}}$ and $2\,\mathrm{km\,s^{-1}}$). These differences are largest in cool giant spectra with line-strengths differing by order 1-$2\,${\AA} with a change of microturbulent velocity from $1\,\mathrm{km\,s^{-1}}$ to $2\,\mathrm{km\,s^{-1}}$. We refer interested readers to section 4.1 of \citealt{Knowles19} for more details. 

 We and other authors have shown that in absolute terms, microturbulence can have a large effect on spectra (\citealt{Conroy2012a}; \citealt{Knowles19}). Therefore, for any absolute application of the model library, it will be important to make a careful consideration of this parameter. Two typical options for this parameter, common in previous libraries, are to compute spectra at fixed microturbulent velocity (e.g. \citealt{Conroy2012a}) or have a varying microturbulent velocity grid dimension (e.g. \citealt{Allende18}). To reduce computation time, but to also incorporate microturbulent velocity values observed in real stars, we have taken a different approach in which spectra are computed with different microturbulent velocity values, depending on the fundamental stellar parameters of $\textrm{T}_{\textrm{eff}}$ ({\small K}) and log g ($\mathrm{cm \,s}^{-2}$).  

We considered three literature representations of how microturbulent velocity (vturb) varies with the physical parameters of stars. These relations were:
\begin{align}
\small{\textrm{vturb} (\mathrm{km\,s^{-1}})=2.478-0.325\hspace{2pt}\textrm{log g}}
\label{APOGEEvturEq}
\end{align}
%\hskip-2.75cm
\begin{multline}
\small{\textrm{vturb} (\mathrm{km\,s^{-1}})=0.871-2.42\times 10^{-4}(\textrm{T}_{\textrm{eff}}-5700)}\\\small{-2.77\times 10^{-7}(\textrm{T}_{\textrm{eff}}-5700)^{2}-0.356(\textrm{log g}-4)}
\label{Thygesen2012Eq}
\end{multline}
\begin{multline}
\small{\textrm{vturb} (\mathrm{km\,s^{-1}})=0.998+3.16\times10^{-4}(\textrm{T}_{\textrm{eff}}-5500 )-0.253(\textrm{log g}-4)}\\\small{-2.86\times10^{-4}(\textrm{T}_{\textrm{eff}}-5500)(\textrm{log g}-4)}\footnotesize{+0.165(\textrm{log g} -4)^2}
\label{DF2016Eq}
\end{multline}
 equation~(\ref{APOGEEvturEq}) was used by APOGEE (\citealt{Holtzman2015}) and was derived using a calibration subsample of red giants, but did not account for any relationship between ${\textrm{T}}_{\textrm{eff}}$ and vturb. Equation ~(\ref{Thygesen2012Eq}) is from \cite{Thygesen2012} using a sample of 82 red giants in the Kepler field. Although this accounted for both effective temperature and surface gravity effects, it was limited to only red giants in a small $\textrm{T}_{\textrm{eff}}$ range ($\approx$4000-5000{\small K}). In the figures we refer to this equation~(\ref{Thygesen2012Eq}) as T12. Equation~(\ref{DF2016Eq}), from \cite{DF2016}, was derived using a sample of cool dwarfs and giants in the Hyades cluster and calibrated to predictions of 3{\small D} models. In the figures below we refer to this equation~(\ref{DF2016Eq}) as DF16.\\\\
In general, based on the observations mentioned above, the behaviour of vturb with $\textrm{T}_{\textrm{eff}}$ and log g follows the following criteria:
\begin{itemize}
	\item{vturb is large ($\approx4\,\mathrm{km\,s^{-1}}$) for high $\textrm{T}_{\textrm{eff}}$($\approx$6000{\small K}) and low log g ($\approx$2) (figures 7 and 9 in \citealt{Gray2001}; figure 1 in \citealt{Montalban2007}). This is larger than values reached by the APOGEE relation and therefore it would be unwise to use that relation for our large parameter space.}
	\item{vturb is smaller ($\ll4\,\mathrm{km\,s^{-1}}$) and can be as small as $<1\,\mathrm{km\,s^{-1}}$ at lower $\textrm{T}_{\textrm{eff}}$ ($\approx$ 5000{\small K}) and high log g ($\approx$ 4.5) (figure 5 in \citealt{Ramirez2013})}
	\item {vturb$\approx$2-$3\,\mathrm{km\,s^{-1}}$ at high $\textrm{T}_{\textrm{eff}}$ ($\approx$7500{\small K}) and high log g ($\approx$4.0) (figures 7 and 9 in \citealt{Gray2001}; figure 5 in \citealt{Niemczura2015}; figure 5 in \citealt{Ramirez2013}). Generally this value is lower than present at high $\textrm{T}_{\textrm{eff}}$ ($\approx$7000{\small K}) and low log g ($\approx$2.5) (figure 1 in \citealt{Montalban2007}), as well as lower than values present at low $\textrm{T}_{\textrm{eff}}$ and low log g (figure 7 of \citealt{Gray2001}).}
	\item{As seen in all the observations considered, giants have higher vturb than dwarfs.}
\end{itemize}
Because our model grids span a wide range of stellar parameter space, it was important to include (at least similar to the sense observed) the trends found in all three of the literature relations (equations~\ref{APOGEEvturEq},~\ref{Thygesen2012Eq} and ~\ref{DF2016Eq}) considered. The DF16 equation was calibrated using a sample of both giant and dwarf stars and included both $\textrm{T}_{\textrm{eff}}$ and log g parameters. Therefore, we used this form of equation~(\ref{DF2016Eq}), but with a slight modification of the cross term, such that:
\begin{multline}
\small{\textrm{vturb ($\mathrm{km\,s^{-1}}$)}=0.998+3.16\times10^{-4}(\textrm{$\textrm{T}_{\textrm{eff}}$}-5500 )-0.253(\textrm{log g}-4)}\\\small{-2\times10^{-4}(\textrm{$\textrm{T}_{\textrm{eff}}$}-5500)(\textrm{log g}-4)}\small{+0.165(\textrm{log g} -4)^2}
\label{DF16ModEq}
\end{multline}
The cross term coefficient was modified from $2.86\times10^{-4}$ to $2\times10^{-4}$ to better follow the trends of equation~(\ref{APOGEEvturEq}) in the parameter range of APOGEE and satisfy the above criteria.

Figure~\ref{fig:DF16_old_vs_new_plus_APOGEE_vs_DF16Mod_vs_T12} (Left Panel) shows the difference between the original DF16 (red lines) and modified DF16Mod (black lines) relations, for different values of $\textrm{T}_{\textrm{eff}}$. For $\textrm{T}_{\textrm{eff}}$=5500{\small K}, the equations are the same, so those two lines overlap. Figure~\ref{fig:DF16_old_vs_new_plus_APOGEE_vs_DF16Mod_vs_T12} (Right Panel) plots our modified equation (black lines) and T12 equation (blue lines) for different values of $\textrm{T}_{\textrm{eff}}$ along with the APOGEE calibration (green line), from equation~(\ref{APOGEEvturEq}). 
\begin{figure}
\begin{center}
 \includegraphics[width=\linewidth, angle=0]{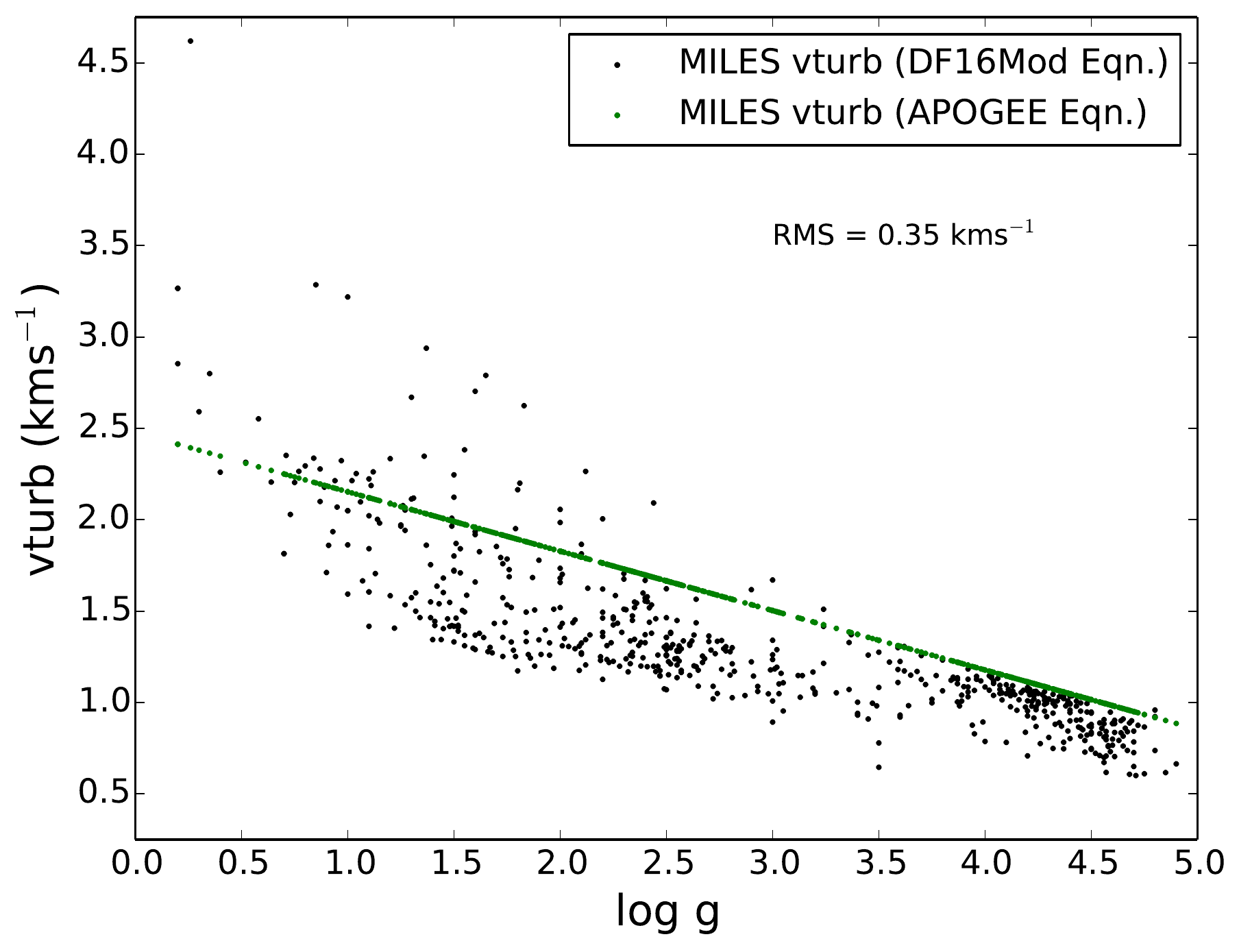}
 \caption[Comparison of DF16Mod and APOGEE microturbulent velocity relations in the \citealt{Cenarro07} MILES parameter range]{Microturbulent velocity as a function of  log g using the modified DF16 equation (DF16Mod black points), and the APOGEE equation (APOGEE green points) for the MILES stars, with stellar parameters from \citealt{Cenarro07}. We also present the RMS scatter between the two estimates. For dwarf stars, DF16Mod agrees well with APOGEE, with larger deviations seen in giant stars.}
 \label{fig:APOGEE_vs_DF16Mod_MILES}
 \end{center}
\end{figure}
 We conclude that it is important to include both effective temperature and surface gravity in the parameterisation, because observations and analyses (e.g. references given above) suggest that trends are present in both. The modified relation~(\ref{DF16ModEq}) approximately follows the trends found in these studies as well as those present in the APOGEE relation~(\ref{APOGEEvturEq}). We used our modified equation (\ref{DF16ModEq}) for $\textrm{T}_{\textrm{eff}}$ from 3500 to 6000{\small K} and for temperatures higher than this we lock the microturbulent velocity to our relation~(\ref{DF16ModEq}) with a fixed $\textrm{T}_{\textrm{eff}}$= 6000{\small K}. To test our parameterisation, we show the difference and RMS scatter between the APOGEE calibration and our relation, for the MILES parameters from \citealt{Cenarro07} in Figure~\ref{fig:APOGEE_vs_DF16Mod_MILES}. This RMS scatter is small compared to the typical values of $1-2\,\mathrm{km\,s^{-1}}$ found for microturbulent velocity in APOGEE (\citealt{Ana2016}).

However, we note that whilst there can be large absolute differences in spectral line-strengths due to microturbulence, we showed in \cite{Knowles19} that effects on the differential application of models were small ($\approx0.02\,${\AA}) compared to typical observational errors on line-strengths ($\approx0.1\,${\AA}). Therefore, for work involving the semi-empirical library, which uses the models only in a differential sense, the choice of microturbulent velocity is not as important as it is for the absolute predictions of models. We still however attempt to match the microturbulent velocity to observations in the generation of theoretical stellar spectra, by using equation~(\ref{DF16ModEq}).

\subsection{New Theoretical Star Grids}
\label{sec:NewGrids}
Due to coverage in log g of the available ODFs, the models were split into three sub-grids, based on ranges in {$\textrm{T}_{\textrm{eff}}$}. All of the models described below were generated using LTE assumptions in both atmosphere and spectral synthesis components.
\subsubsection{3500-6000{\small{K}} Grid}
For the lowest temperature grid, models were computed with the following parameter steps, such that:
\begin{itemize}
    \item {$\textrm{T}_{\textrm{eff}}$=3500{\small K} to 6000{\small{K}} in steps of 250{\small K}}
    \item {log g=0 to 5 in steps of 0.5 dex}
    \item {[M/H]=$-$2.5 to +0.5 in steps of 0.5 dex}
    \item {[$\alpha$/M]=$-$0.25 to +0.75 in steps of 0.25 dex. We note here that we are making an assumption that these elements increase in lockstep, which is not exactly true in the Milky Way (e.g. \citealt{Bensby2014}; \citealt{Zasowski19})}
    \item{[C/M]=$-$0.25 to 0.25 in steps of 0.25 dex}
\end{itemize}
Thus, the number of models computed in this grid is
\begin{center}
Number of Models =  $\textrm{T}_{\textrm{eff}}$ steps x log g steps x Element Variations \\
=N($\textrm{T}_{\textrm{eff}}$) x N(log g) x N([M/H]) x N([$\alpha$/M]) x N([C/M])\\ 
=11 x 11 x 7 x  5 x 3  = 12705 models
\end{center}
For these 12705 models, seven models were missing ODFs or did not converge. In order to maintain regularity of the grid, the missing models were computed using a linear interpolation of models in the nearest available grid points. These seven models were all at the lowest $\textrm{T}_{\textrm{eff}}$ (3500{\small K}), high surface gravity (log g=4.0, 4.5 5.0), low metallicity ([M/H]=$-$1.5 or 2.0) and at high $\alpha$ abundance ([$\alpha$/M]=0.75)) points. The parameters of these seven stars are specified in \cite{Knowles19_Thesis} (section 3.3.2).
\begin{figure}
 \begin{center}
 \includegraphics[width=82mm,
 angle=0]{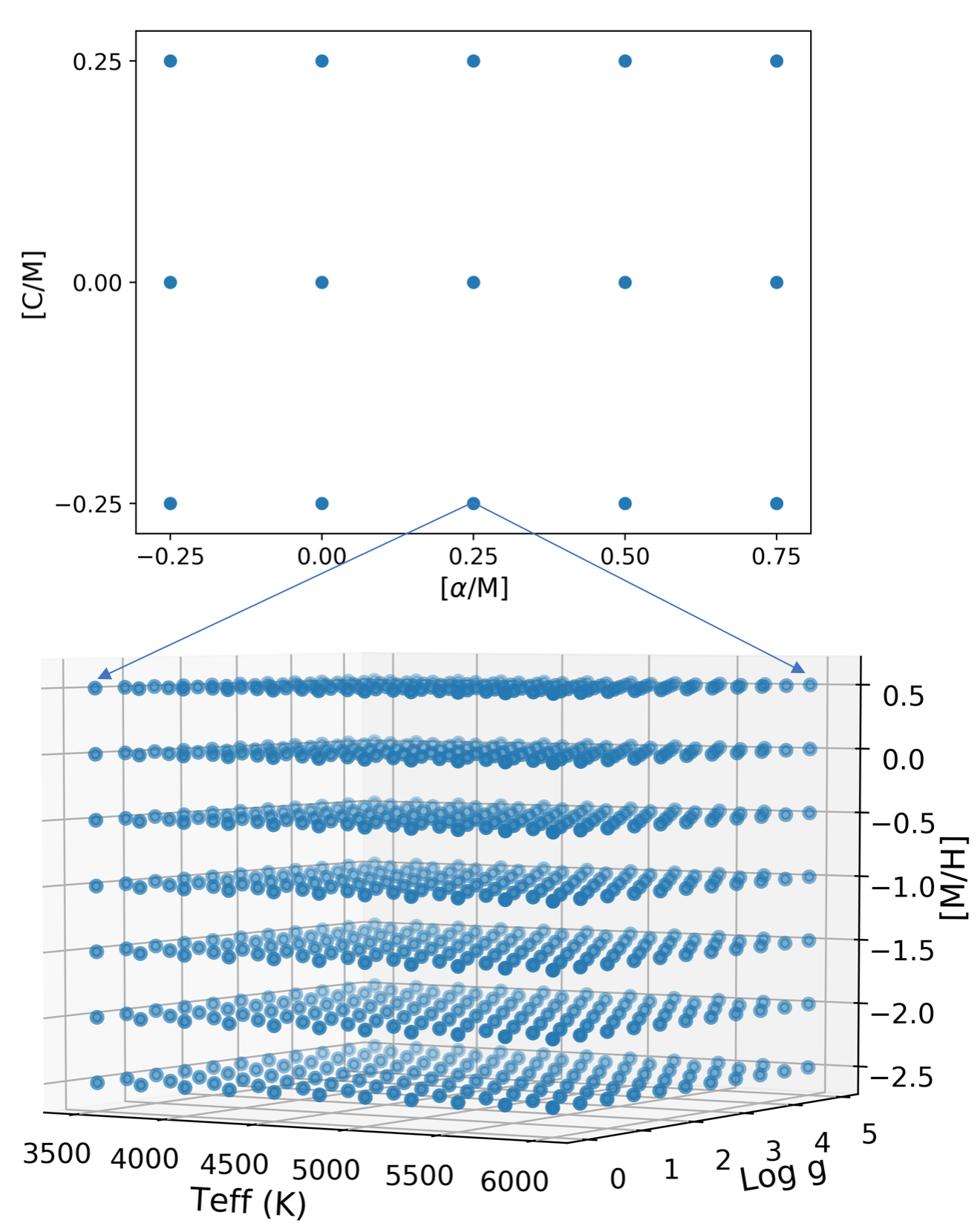}
  \caption{Top Panel: Abundance pattern coverage in the [C/M] vs [$\alpha$/M] plane. Bottom Panel: 3{\small D} stellar parameter coverage of 3500-6000{\small K} grid. Each point in the [C/M] vs [$\alpha$/M] plane represents 11 x 11 x 7 = 847 models in this lowest $\textrm{T}_{\textrm{eff}}$ grid.}
 \label{fig:TheoreticalGrid_Coverage}
 \end{center}
\end{figure}

For illustration, the parameter coverage of the lowest effective temperature grid is presented in Figure~\ref{fig:TheoreticalGrid_Coverage}.
\begin{figure*}
\begin{center}
\includegraphics[width=\linewidth, angle=0]{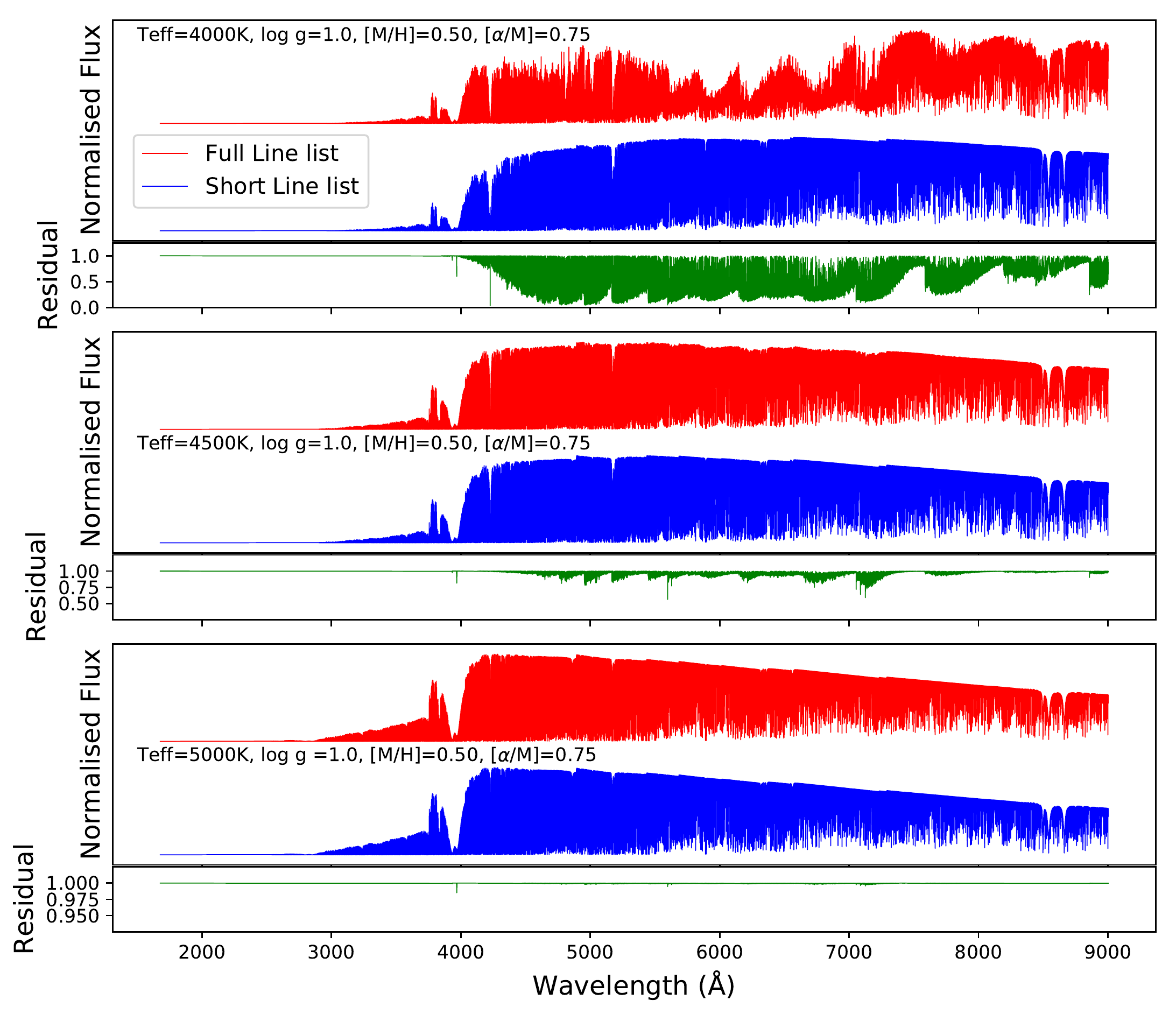}
\caption[Effect of removing TiO lines at different effective temperatures in the high resolution theoretical stellar library]{Effect of removing TiO lines from molecular line list at different temperatures for the fixed binning, high-resolution library described in Section 2.6. The red and blue spectra represent stars with the TiO line list included and removed for each temperature, respectively. Fluxes are normalised to the maximum flux value of each spectrum. The green line represents the residual obtained from a division of Full line list and Short line list spectra. Differences in the top panel ($\textrm{T}_{\textrm{eff}}$=4000K) are seen in locations known to be affected by TiO absorption (see  5a of \citealt{Kirkpatrick91}; figure 1 of \citealt{Plez98}; figure 1 of \citealt{Allard00}).}
\label{fig:Linelist_Test}
\end{center}
\end{figure*}
To help minimize the number of models, we split our higher temperature models into two sub-grids. We have a grid of models from 6250-8000{\small K} and a grid from 8250-10000{\small K}. The upper limit of these temperatures was chosen to cover regions of the existing MILES library where stars that contain the most information regarding abundance patterns exist.
The ODFs and model atmospheres available also make cuts to surface gravity at the higher temperatures, which have increasing radiation pressure and therefore the lowest surface gravity models become unstable (e.g. see figure 2 of \citealt{Mezaros2012}). Thus, the number of models for our higher $\textrm{T}_{\textrm{eff}}$ sub-grids are described in Sections~\ref{sec:62508000Grid} and ~\ref{sec:825010000Grid}.
\subsubsection{6250-8000{\small K} Grid}
\label{sec:62508000Grid}
\begin{itemize}
\item{$\textrm{T}_{\textrm{eff}}$=6250{\small K} to 8000{\small K}, in steps of 250{\small K}}
\item{log g=1 to 5, in steps of 0.5 dex}
\item{[M/H]=$-$2.5 to +0.5, in steps of 0.5 dex}
\item {$[\alpha/\textrm{M}]$=$-$0.25 to +0.75, in steps of 0.25 dex}
\item{[C/M]=$-$0.25 to +0.25, in steps of 0.25 dex}
\end{itemize}
Thus, the number of models computed in the 6250-8000{\small K} grid is
\begin{center}
N($\textrm{T}_{\textrm{eff}}$) x N(log g) x N([M/H]) x N([$\alpha$/M]) x N([C/M])\\
 8 x 9 x 7 x  5 x 3  = 7560 models
\end{center}
To avoid excessive computation times, careful consideration of the number of spectra, the wavelength coverage, linelists used and number of abundance steps was necessary. A method to decrease computation time is to reduce the number of input atomic and molecular transitions. For $\textrm{T}_{\textrm{eff}}$ above 6000{\small K}, we removed a significant molecular contributor to the linelists, TiO, which is prevalent in stellar spectra at low temperatures, however at higher temperatures absorption features become weak. TiO band strengths are particularly used in unresolved stellar population analysis as Initial Mass Function (IMF) probes (e.g. TiO$_2$ defined in \citealt{Trager98}). For example, \cite{LaBarbera16} use TiO index measurements to investigate the radial variations of the IMF in ETGs. This index strength increases as effective temperature decreases and therefore the IMF sensitivity arises from the ratio of low mass (low effective temperature) to high mass stars on the main sequence (\citealt{Fontanot18}). Figure~\ref{fig:Linelist_Test} shows an example of the effect of removing TiO transitions from our models at various temperatures. As expected, TiO bands are extremely prevalent in the lowest $\textrm{T}_{\textrm{eff}}$ spectrum and differences in the grid between higher temperature models are very small.

\subsubsection{8250-10000{\small K} Grid}
\label{sec:825010000Grid}
\begin{itemize}
\item{$\textrm{T}_{\textrm{eff}}$=8250{\small K} to 10000{\small K} in steps of 250{\small K}}
\item{log g=2 to 5, in steps of 0.5 dex}
\item{[M/H]=$-$2.5 to +0.5, in steps of 0.5 dex}
\item {$[\alpha/\textrm{M}]$=$-$0.25 to +0.75, in steps of 0.25 dex}
\item{[C/M]=$-$0.25 to 0.25, in steps of 0.25 dex}
\end{itemize}
Thus, the number of models computed in the 8250-10000{\small K} grid is
\begin{center}
N($\textrm{T}_{\textrm{eff}}$) x N(log g) x N([M/H]) x N([$\alpha$/M]) x N([C/M]) =\\
 8 x 7 x 7 x  5 x 3  = 5880 models
\end{center}
No models in the two higher $\textrm{T}_{\textrm{eff}}$ grids had missing ODFs or convergence issues.
\subsubsection{[Ca/Fe]=0 Grid}
We also compute a small model grid with [Ca/Fe]=0.0 to match results of integrated light studies of ETGs in which calcium was found to track iron-peak elements (\citealt{Vazdekis97}; \citealt{Trager98}; \citealt{Thomas03Ca}; \citealt{Schiavon2007}; \citealt{Johansson12}; \citealt{Conroy2014}).
\begin{itemize}
\item{$\textrm{T}_{\textrm{eff}}$=3500{\small K} to 6000{\small K}, in steps of 250{\small K}}
\item{log g=0 to 5, in steps of 0.5 dex}
\item{[M/H]=$-$2.5 to +0.5, in steps of 0.5 dex}
\item{[$\alpha$/M]=0.25, where $\alpha$ is O, Ne, Mg, Si, S and Ti}
\item{[C/M]=0.25 - as was found by \cite{Conroy2014}}
\end{itemize}
Thus, the number of models computed in the [Ca/Fe]=0.0 grid is
\begin{center}
N($\textrm{T}_{\textrm{eff}}$) x N(log g) x N([M/H]) x N([$\alpha$/M]) x N([C/M]) =\\
 11 x 11 x 7 x  1 x 1  = 847 models
\end{center}
\subsection{Processing}
We now describe methods and procedures of processing raw spectra from ASS$\epsilon$T into three different resolution libraries: a high resolution library in which there is a fixed resolving power (R=$\lambda$/d$\lambda$ - based on equation {\ref{ASSETbinEq}}) within a spectrum but each spectrum has a different resolving power and sampling, a high resolution theoretical library in which all spectra are binned to a common wavelength range and sampling, and a MILES resolution theoretical library used in the differential correction process.

ASS$\epsilon$T generates a spectrum in wavelength (in \AA) and flux density measured at the stellar surface (in erg/s/$\mathrm{cm}^{2}$/\AA). Spectra are computed at fixed resolving power, resulting in a sampling that is constant in $d(\log_{10}\lambda)$ but increasing $d\lambda$  for increasing $\lambda$. 

As default, ASS$\epsilon$T samples the spectrum based on the formula:
\begin{equation}
    \textrm{d}(\log_{10}\lambda)=0.3\sqrt{(v_{Micro}^{2}+v_{TM}^{2})},
    \label{ASSETbinEq}
\end{equation}
where $v_{Micro}$ is the microturbulent velocity and $v_{TM}$ is the thermal Doppler width computed in ASS$\epsilon$T at the coolest layer of the atmosphere. This formula ensures the sampling of at least three wavelength points for the expected line width of the spectrum, but means that every spectrum was computed at different sampling. This is the first theoretical library generated, in which each spectrum has a unique sampling and fixed resolving power based on equation~(\ref{ASSETbinEq}).

The IRAF task `dispcor' was then used to resample the spectra, with fifth order polynomial interpolation, to a common start and end wavelength as well as number of wavelength points. Flux density was conserved throughout the resampling process.  The common sampling was taken as the largest sampling value of all the spectra generated. This resulted in a final, high resolution library consisting of spectra with $\lambda_{\textrm{start}}=1677.10\,${\AA}, d$\lambda=0.05\,${\AA} and number of wavelengths points, $n_{\lambda}$=146497. This is the second library mentioned above, in which high resolution theoretical spectra are produced with all spectra at a common wavelength range, sampling and resolution.

To create synthetic spectra that replicate existing MILES stars, with which differential corrections will be performed, the theoretical library was matched to the existing MILES empirical library in terms of wavelength range, sampling and resolution. IDL routines were used to smooth and rebin\footnote{IDL routines were from https://ascl.net/1708.005, plus our own IDL routine for rebinning by summing and renormalising to relative flux density.},  the fixed sampling, high resolution theoretical library to match the MILES empirical spectra, resulting in a third library with a wavelength range, sampling and resolution of $3540.5-7409.6\,${\AA}, $0.9\,${\AA} and $2.5\,${\AA} respectively. Models of existing MILES stars and MILES stars with different abundance patterns are created via interpolation in these MILES specific theoretical libraries, as described in Section~\ref{sec:sMILESstars}.

\subsection{Theoretical Library Summary}
In summary, three grids of theoretical stellar spectra were computed, covering different $\textrm{T}_{\textrm{eff}}$ ranges. The first grid consisted of spectra covering effective temperatures from 3500 to 6000{\small K}, surface gravity from 0 to 5 dex and metallicities ([M/H]) that covered a large proportion of the MILES empirical library. Models in this first grid were computed with a microturbulent velocity according to equation~\ref{DF16ModEq}. The second grid, was computed with the same coverage in metallicity as the first grid, but with an effective temperature coverage from 6250 to 8000{\small K} and coverage in surface gravity from 1 to 5 dex, to avoid unstable model atmospheres caused by radiation pressure instabilities.  The third grid was also computed with the same coverage in metallicity as the first, but with an effective temperature coverage from 8250 to 10000{\small K} and surface gravity coverage from 2 to 5 to also avoid unstable model atmospheres caused by radiation pressure instabilities. Both the 6250-8000{\small K} and 8250-10000{\small K} grids were computed with a reduced linelist in which TiO was removed, in order to shorten computation times. Models in the second and third grid were computed with a microturbulent velocity according to  equation~\ref{DF16ModEq}, with $\textrm{T}_{\textrm{eff}}$ fixed at 6000{\small K}. All three grids were computed with [$\alpha$/M] variations that cover a range of $\alpha$ abundance variation as observed in external systems such as ETGs galaxies and dSphs, as well as [C/M] variations in a range that covered observations in previous integrated light studies (e.g. \citealt{Conroy2014}; \citealt{WortheyTangServen2014}). Example sequences of theoretical spectra for the parameters of $\textrm{T}_{\textrm{eff}}$, [M/H], [$\alpha$/M] and [C/M] are presented in the supplementary data provided.

Each of these restricted temperature grids exists at three different resolution and sampling values. The first library (collection of three temperature grids) is one in which each spectrum has a unique sampling and resolving power based on equation~(\ref{ASSETbinEq}). 

The second library consists of spectra with a common wavelength range and sampling, such that $\lambda_{\textrm{start}}=1677.10\,${\AA}, d$\lambda=0.05\,${\AA} and $n_{\lambda}$=146497. This fixed binning, high-resolution library is publicly available to download at \url{http://uclandata.uclan.ac.uk/178/}. This library consists of three grids; a low temperature grid ($\textrm{T}_{\textrm{eff}}$=3500-6000{\small K}), an intermediate temperature grid ($\textrm{T}_{\textrm{eff}}$=6000-8000{\small K}) and a high temperature grid ($\textrm{T}_{\textrm{eff}}$=8000-10000{\small K}). The higher two grids include repeats of the highest temperature spectra from the grid below, to maintain continuous coverage in $\textrm{T}_{\textrm{eff}}$. 

Finally, a MILES-specific theoretical library exists, with spectra smoothed and resampled to match the current MILES empirical library. The result is a medium resolution library with spectra that have a wavelength range, sampling and resolution (FWHM) of $3540.5-7409.6\,${\AA}, $0.9\,${\AA} and $2.5\,${\AA} respectively. This library and its predictions will be used in the later sections of this work to create semi-empirical stellar spectra.

We refer to our computed models as the ATK set in later sections of this work.

\section{Testing Theoretical Library}
\label{sec:ModelTesting}
 We now make comparisons between our models and other published libraries of theoretical stellar spectra.
\begin{table*}
\caption{Theoretical star spectra compared for our current models and those published in \protect\cite{Allende18}. The vturb values are those used (see Section~\ref{sec:Microturbulence}) and the Allende Prieto models were interpolated to those values. We have tested a giant (G) and two dwarf (D$_1$, D$_2$) stars. Also shown is RMS scatter about the 1:1 agreement line between the differential predictions ([$\alpha$/M]=0.25/[$\alpha$/M]=0.0) of our models and Allende Prieto models. RMS is calculated for the ratio of our (ATK) and Allende Prieto (CAP) sets of differential predictions of spectra with different abundance patterns.}
\begin{tabular}{ccc}
\hline
Star Type              & RMS ($\lambda$<$3000\,${\AA}) & RMS ($\lambda$>$3000\,${\AA})\\
\hline
$\textrm{[M/H]}=0.0$&&\\
G (T$_\textrm{eff}$=4000K, log g=2.0, vturb=$1.09\,\mathrm{km\,s^{-1}}$)   & $\mathrm{1.33\times10^{-2}}$ & $\mathrm{9.00\times10^{-4}}$ \\
D$_1$ (T$_\textrm{eff}$=4000K, log g=4.0,vturb=$0.524\,\mathrm{km\,s^{-1}}$)  & $\mathrm{6.75\times10^{-3}}$ & $\mathrm{7.34\times10^{-4}}$  \\
D$_2$ (T$_\textrm{eff}$=5500K, log g=4.0,vturb=$0.998\,\mathrm{km\,s^{-1}}$)     & $\mathrm{2.68\times10^{-3}}$& $\mathrm{2.57\times10^{-4}}$ \\ \\
$\textrm{[M/H]}=-1.0$&&\\
G (T$_\textrm{eff}$=4000K, log g=2.0, vturb=$1.09\,\mathrm{km\,s^{-1}}$)  &$\mathrm{1.20\times10^{-2}}$ & $\mathrm{6.35\times10^{-4}}$  \\
D$_1$ (T$_\textrm{eff}$=4000K, log g=4.0,vturb=$0.524\,\mathrm{km\,s^{-1}}$)  &$\mathrm{4.12\times10^{-3}}$ & $\mathrm{3.96\times10^{-4}}$  \\
D$_2$ (T$_\textrm{eff}$=5500K, log g=4.0,vturb=$0.998\,\mathrm{km\,s^{-1}}$)  & $\mathrm{1.70\times10^{-3}}$ & $\mathrm{1.19\times10^{-4}}$\\ \hline
\end{tabular}
\label{ATK_vs_CAP_RMS}
\end{table*}
\subsection{Comparison to Allende Prieto models}
\label{sec:ATKvsCAP}
To check the accuracy of our theoretical spectra, we first compare to the library of \protect\cite{Allende18}, which covers a wide range of star types, metallicities, [$\alpha$/Fe] and microturbulent velocities. We refer to these models as the CAP set throughout this work. We focus on the differential abundance pattern predictions of both model sets. The abundance pattern prediction is taken as a ratio of an $\alpha$-enhanced ([$\alpha$/M]=0.25) and  solar abundance pattern ([$\alpha$/M]=0.0 star. \protect\cite{Allende18} models were interpolated in microturbulent velocity, using a quadratic B\'ezier function, to match stars in our library. Options for interpolations within model grids are discussed in Appendix~\ref{sec:InterpChoice}. Table~\ref{ATK_vs_CAP_RMS} lists the star types compared and their parameters.

\begin{figure*}
\begin{center}
\includegraphics[width=152mm, angle=0]{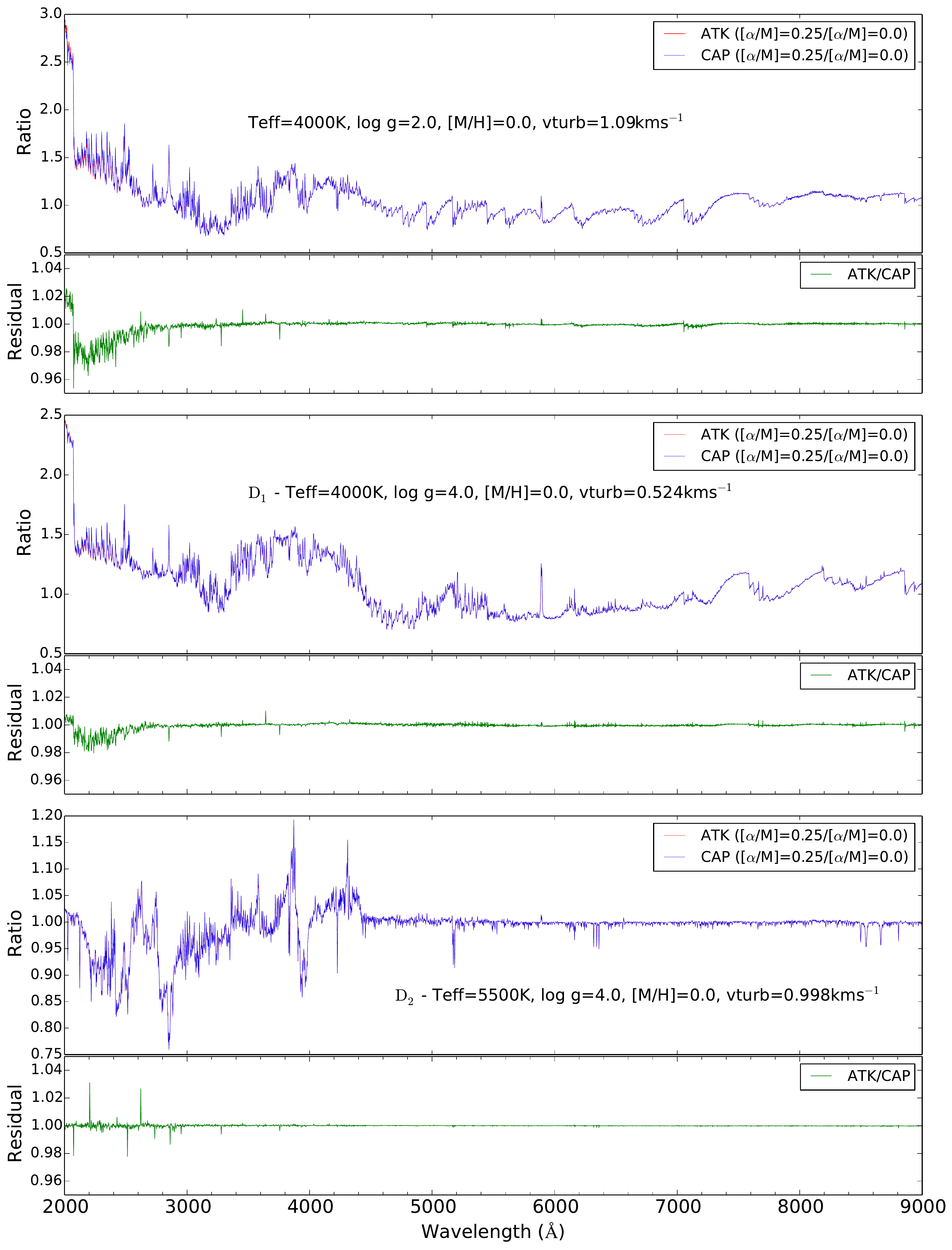}
\caption{Comparisons of enhanced-over-base star spectra for stars in our theoretical library, labelled ATK, versus stars from \citealt{Allende18} (interpolated in vturb), labelled CAP. The top plot in each block shows spectra for [$\alpha$/M]=+0.25 divided by [$\alpha$/M]=0.0. The lower plot in each block shows the division of these two ratios (ATK/CAP). Blocks show comparisons for a giant star (upper) and two dwarf stars (middle and lower), all at solar metallicity, with parameters detailed in Table~\ref{ATK_vs_CAP_RMS}.}
\label{fig:ATK_vs_CAP_MH0.0}
\end{center}
\end{figure*}

\begin{figure}
\begin{center}
\includegraphics[width=86mm, angle=0]{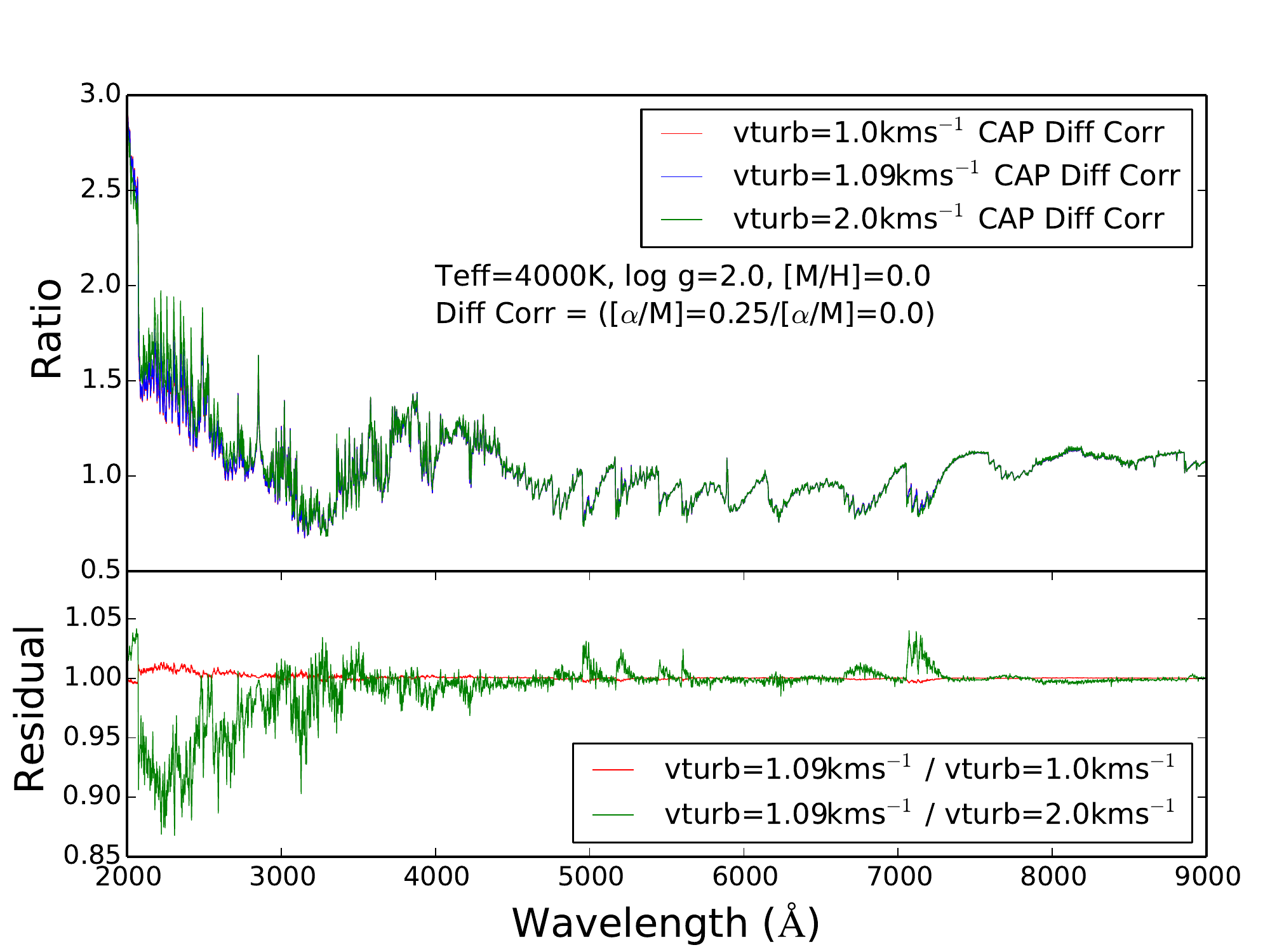}
\caption{Effect of vturb on CAP model abundance pattern predictions. The effect of vturb on the differential correction is strongest in the UV, with large differences present at the shortest wavelengths between vturb=$1.09\,\mathrm{km\,s^{-1}}$ and vturb=$2\,\mathrm{km\,s^{-1}}$.}
\label{CAP_vturb_Test}
\end{center}
\end{figure}

 Both model sets were degraded to a spectral resolution of $2.5\,${\AA} FWHM and resampled to $0.3\,${\AA} bins, in order to compare spectra across the full wavelength range available (2000 to $9000\,${\AA}). Figure~\ref{fig:ATK_vs_CAP_MH0.0} shows the comparisons between abundance pattern predictions, at solar metallicity. In all cases the difference between predictions above $3000\,${\AA} is small, with RMS values about the 1:1 model agreement of 0.000900, 0.000734, 0.000257 for the cool giant, coolest dwarf (D$_1$) and cool dwarf (D$_2$) star respectively. The largest deviations are found below $3000\,${\AA} with RMS values of 0.0134, 0.00675 and 0.00268 for giant, D$_1$ and D$_2$ star respectively. Similar results were found for the same analysis at [M/H]=$-$1.0, with RMS values summarised in Table~\ref{ATK_vs_CAP_RMS}.

As both \cite{Allende18} and the current set of models use similar methods in the computation of stellar spectra, it is important to show that they produce very similar predictions. The exception to this is found in the UV, where differences between the models are larger. These differences may be due to a combination of four effects. Firstly, the fine grids of \cite{Allende18} (described in section 2.4 of that work) use cubic interpolations of the opacity as a function of density and temperature to reduce computation times, whereas our models use the 'ONE-MOD' mode in ASS$\epsilon$T to compute the opacity for each model at every depth. This difference is expected to be largest in the UV, where more metal lines are present. Secondly, the models of \cite{Allende18} were computed with the outermost layers of the stellar atmospheres removed, which are less reliable for stars with T$_\textrm{eff}$<5000{\small K} (\citealt{Mezaros2012}). Thirdly, the method of microturbulent velocity handling in model sets may also cause small differences in the predictions. Lastly is the inclusion of neon in the $\alpha$-elements of our models. The opacity treatment and outer layer removal in the calculations are expected to be the dominant effects and can cause flux differences on the order of a few percent in the UV, in agreement with the values shown in Figure~\ref{fig:ATK_vs_CAP_MH0.0}.

To highlight the impact of vturb on the UV CAP model predictions, we plot a comparison between the differential corrections predicted with vturb=1, 1.09 and $2\,\mathrm{km\,s^{-1}}$ in Figure~\ref{CAP_vturb_Test}. The vturb=1 and $2\,\mathrm{km\,s^{-1}}$ spectra are existing grid points in the published grids. vturb=$1.09\,\mathrm{km\,s^{-1}}$ spectra were generated using interpolations within the ATK grid. As shown, the effect of microturbulence is largest at UV wavelengths, with significant differences found. In our previous work (\citealt{Knowles19}) we showed that uncertainties in vturb can have large effects on the absolute predictions of spectral features, but only small effects on differential corrections in the MILES wavelength range. Our current work shows that differential corrections are more strongly affected by vturb below $\sim$ $3500\,${\AA}. Also from \cite{Knowles19}, we show that spectral models are more similar to each other than they are to real stars. We illustrate some comparisons between our models and real stars in Section~\ref{sec:TestingObs} to show where they agree well and where work is most needed.

In summary, comparisons between ATK and CAP predictions of abundance pattern effects have shown that they agree well in the MILES wavelength range, which is important in the generation of semi-empirical stars described later. Small differences between model predictions are found for wavelengths below $\sim$ $3000\,${\AA}, which may be attributed to differences in the method of opacity treatment and interpolation effects when generating CAP models with same microturbulent velocity as ATK models. Differences in microturbulence can have large effects (up to $\sim$10 percent) on model abundance pattern predictions below $\sim$ $3500\,${\AA}. Neon inclusion in the $\alpha$-elements of ATK models may also create small differences between differential corrections. Further work is required to fully assess these small differences between models in the UV and is beyond the scope of this current work.

\subsection{Comparison to PHOENIX Models}
\label{ATK_vs_PHOENIX}
\begin{table*}
	\centering
\caption{Methods used in the generation of theoretical stellar spectra, for our model grid (ATK) and PHOENIX model grid (\citealt{Husser2013}).}
	\begin{tabular}{cp{20mm}p{15mm}p{45mm}p{15mm}p{25mm}p{15mm}}
		\hline
		Model & Atmosphere Code & Spectrum Code & Equation of State & vturb & Solar Abundance Reference & $\alpha$-elements\\
		\hline
		ATK  & ATLAS9, LTE, Plane-parallel (\citealt{Kurucz1993}) & ASS$\epsilon$T & Synspec (\citealt{Hubeny2017}, for the first 99 atoms and 338 molecules (\citealt{Tsuji64,Tsuji73,Tsuji76}, with partition functions from \cite{Irwin81} and updates. & Equation~\ref{DF16ModEq} in Section~\ref{sec:Microturbulence} & \cite{Asplund2005} & O, Ne, Mg, Si, S, Ca, Ti \\
		PHOENIX  & PHOENIX, LTE, Spherical based on \cite{Hauschildt99} & PHOENIX & Astrophysical Chemical Equilibrium Solver (ACES, see \citealt{Husser2013}) for 839 species (84 elements, 289 ions, 249 molecules, 217 condensates.) & Section 2.3.3 and Equation 7 of \cite{Husser2013} & \cite{Asplund2009} & O, Ne, Mg, Si, S, Ar, Ca, Ti \\
		\hline
	\end{tabular}
	\label{ATK_PHOENIX}
\end{table*}

 We now compare to another up-to-date and widely-used theoretical stellar spectral library of \cite{Husser2013}, hereafter referred to as the PHOENIX library. Again, we test the relative changes due to variations in atmospheric abundances, rather than focusing on the absolute predictions, which are already known to have limitations as described in Section~\ref{sec:ModelSpectra}. The PHOENIX library consists of high resolution stellar spectra that cover a wide range of stellar parameters and [$\alpha$/Fe] abundances, making it an ideal set to compare to our models. PHOENIX spectra were generated from an updated version of the PHOENIX stellar atmosphere code, described in \cite{Husser2013} and references therein. We use the publicly available distribution of the PHOENIX library\footnote{\url{http://phoenix.astro.physik.uni-goettingen.de/}} in the comparisons. We compare our models to the medium resolution (FWHM=$1\,${\AA}) version of the PHOENIX library. There are several differences between the computation methods of our models and the PHOENIX grid that we summarise in Table~\ref{ATK_PHOENIX}.

 Both sets of models use the same definitions of [Fe/H] and [$\alpha$/Fe] as described in Section~\ref{sec:CompMethod}, in that the total metallicity (Z) is not conserved when [$\alpha$/Fe] is changed. These definitions mean that [$\alpha$/Fe] and [$\alpha$/M], with M defined in equation~(\ref{MHDefEq}), are equivalent and can be used interchangeably. However, the solar abundances adopted in PHOENIX are from \cite{Asplund2009} compared to the values of \cite{Asplund2005} adopted in ATK models. We compare differential predictions of atmospheric $\alpha$ abundance variations, in the same way as Section~\ref{sec:ATKvsCAP}. This is done for two representative star types; a giant (Teff=4500{\small K}, log g=1.5, [M/H]=0.0) and dwarf (Teff=5500{\small K}, log g=4.0, [M/H]=0.0) star. The microturbulent velocity in both ATK and PHOENIX models are very similar, to minimise any differences due to this parameter.

We generate models to match the PHOENIX [$\alpha$/Fe] enhancement of 0.20 using a quadratic interpolation within FER\reflectbox{R}E\footnote{Publicly available at \url{https://github.com/callendeprieto/ferre}.} (\citealt{Allende2006}). We test the predictions of how an [$\alpha$/Fe] change affects spectra, through ratios of enhanced and solar abundance pattern stars ([$\alpha$/Fe]=0.2/[$\alpha$/Fe]=0.0) for both model sets. ATK models were degraded to $1\,${\AA} resolution and resampled to match the PHOENIX spectra. PHOENIX spectra were also converted to air wavelengths to match the ATK models, using the conversion described in section 2.4 (their equations 8, 9 and 10) of \cite{Husser2013}, which is based on \cite{Ciddor96}.

\begin{figure*}
\begin{center}
\includegraphics[width=154mm,angle=0]{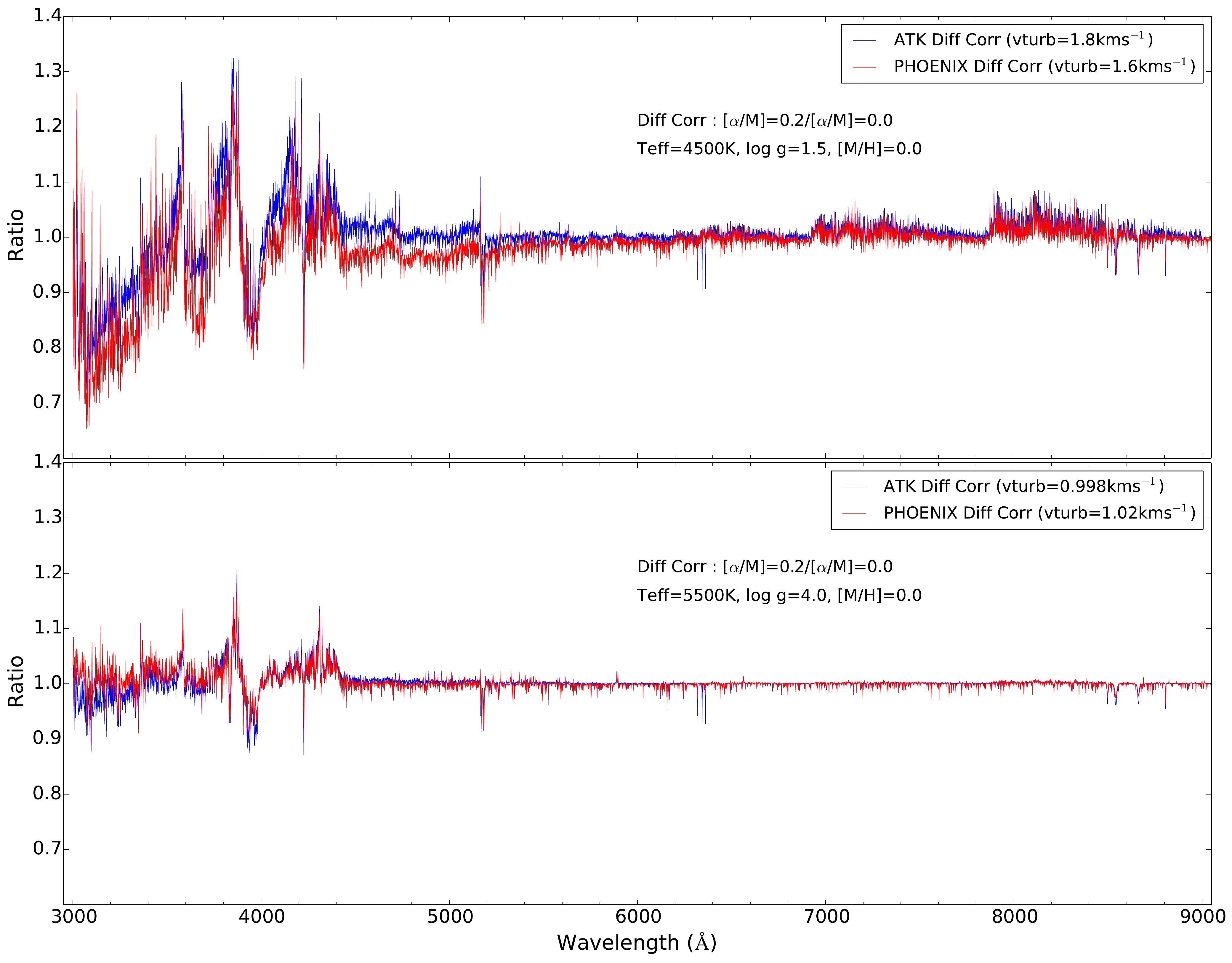}
\caption{Comparison between predicted differential corrections of ATK and PHOENIX models. Both sets of models are smoothed to $1\,${\AA} FWHM and sampled in air wavelengths. Blue and red lines represent ATK and PHOENIX model predictions, respectively. Top panel: A comparison of a giant star differential correction. Bottom panel: A comparison of a dwarf star differential correction.}
\label{fig:ATK_vs_PHOENIX_Spec}
\end{center}
\end{figure*}

\begin{figure*}
\begin{center}
\includegraphics[width=154mm,angle=0]{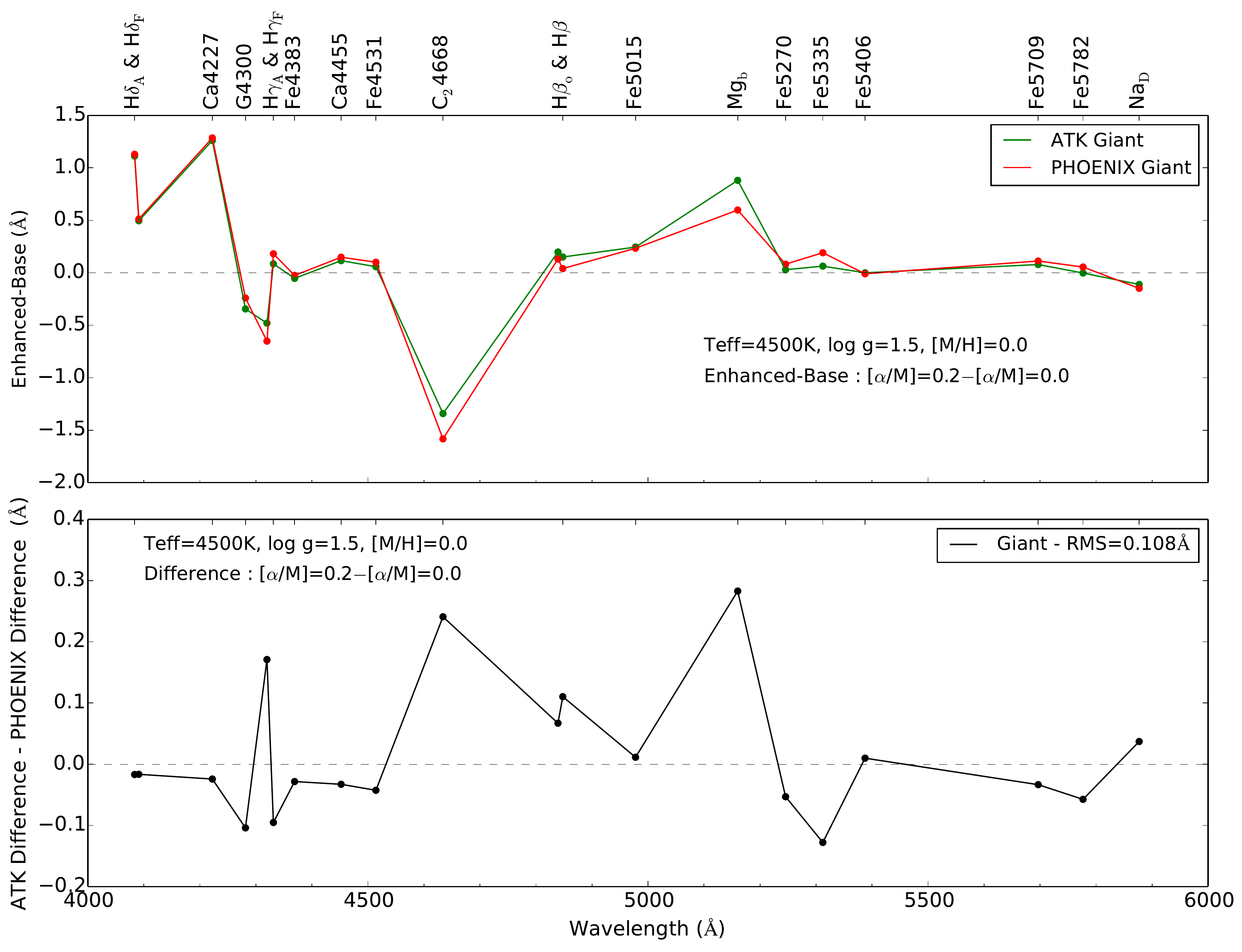}
\caption{Comparison between ATK and PHOENIX giant star model predictions of the change of Lick indices due to an atmospheric enhancement of $\alpha$-elements. This is done for Lick indices that are measured in {\AA}, including H$\beta_{\textrm{o}}$ from \protect\cite{Cervantes09}. Top panel: Change of Lick indices due to an $\alpha$ enhancement of 0.2. Bottom panel: The difference between the changes in ATK and PHOENIX models. The 1:1 agreement between model predictions is plotted as a dashed horizontal line. Lick indices are labelled for illustration. Note that in this comparison [Fe/H] is kept constant.}
\label{fig:ATK_vs_PHOENIX_Lick_Giant}
\end{center}
\end{figure*}

Figure~\ref{fig:ATK_vs_PHOENIX_Spec} shows the comparison of model predictions. For both star types, the general shape of both ATK and PHOENIX differential predictions are similar. However, there are offsets that are generally larger at shorter wavelengths, where metal lines are are more prevalent. For the giant star, ATK models predict a smaller differential correction (i.e. a smaller reduction in flux due to an atmospheric $\alpha$-enhancements), with larger offsets between ATK and PHOENIX models seen below $\sim5000\,${\AA}. For the dwarf star, the opposite behaviour is found, with ATK models predicting a larger reduction in flux at the shortest wavelength values. Above $\sim5000\,${\AA} in both star types, there is a reasonable agreement between models, with the exception of three features at $\sim$6318, 6343 and $6362\,${\AA}, in which ATK models predict a much larger change than the PHOENIX models. These are known calcium-sensitive lines and are found to be Ca auto-ionization lines, observed as broad lines in late-type stars (\citealt{Culver67}; \citealt{Barbuy15}). On closer inspection, PHOENIX models include the first two of these features but appear to be missing the reddest line. Given that the spectral models differ in every component of the computation, from the atmosphere and radiative transfer modelling through to the equation of state, line lists and even the reference solar abundance, finding the main cause of the offsets is a difficult task and beyond the scope of this work. It is likely that every difference in the calculations contributes to the these offsets. Despite the differences in methodology, we find that generally, ATK models predict a differential correction of similar shape and magnitude to PHOENIX models across the full wavelength range tested.

In Figure~\ref{fig:ATK_vs_PHOENIX_Lick_Giant} we investigate how Lick indices (\citealt{Worthey94}; \citealt{Worthey97}; \citealt{Trager98}) change for an [$\alpha$/Fe] enhancement in ATK and PHOENIX models, for the giant star in Figure~\ref{fig:ATK_vs_PHOENIX_Spec}. This is performed for the standard Lick indices that are measured in {\AA}, including H$\beta_{\textrm{o}}$ defined in \cite{Cervantes09}. In Figure~\ref{fig:ATK_vs_PHOENIX_Lick_Giant} the change is now represented as a subtraction, rather than a ratio as in Figure~\ref{fig:ATK_vs_PHOENIX_Spec}. The top panel in Figure~\ref{fig:ATK_vs_PHOENIX_Lick_Giant} shows a direct comparison between ATK and PHOENIX model predictions of changes in Lick indices and the bottom panel shows the difference of predicted changes between models. In the bottom panel we also show the RMS scatter about the 1:1 agreement line (dotted horizontal line). The model predictions are similar, with an RMS value ($0.108\,${\AA}) comparable to typical observational uncertainties in Lick line strengths ($\sim$ 0.1 dex - e.g. see Table 2 of \citealt{Sansom2013}). The analysis is also performed for the dwarf star in Figure~\ref{fig:ATK_vs_PHOENIX_Spec} and an RMS value of $0.0718\,${\AA} is found. Larger differences between model predictions are seen for a few indices, including C$_2$4668 and Mg$_b$, for both giant and dwarf stars. One significant difference between models is the inclusion of spherical geometry in the atmospheric structure of PHOENIX models, compared to the 1{\small D} calculations of ATK. \cite{Bergemann12Fe,Bergemann17Mg} show that low-excitation FeI and Mg lines are sensitive to atmospheric structure and that the effect of NLTE on line strengths and abundance predictions can vary depending on whether the underlying atmosphere is calculated in 1{\small D} or 3{\small D}. In these works they find that for a giant star, with T$_\textrm{eff}$ and log g values similar to those tested here, the 1{\small D} LTE models predict a slightly larger metallicity and magnesium abundance than 3{\small D} LTE models. We note however that in these works, the atmosphere calculations are not fully 3{\small D} and are computed through time and spatial averages (<3{\small D}>) of full hydrodynamical simulations. Systematic errors in abundances determinations were found for <3{\small D}> LTE models in these works. In this work, we find that for Fe5270 and Fe5335 ATK models predict smaller Lick indices than PHOENIX in both the solar and $\alpha$-enhanced giant star. For Mg$_\textrm{b}$, we find that the Lick indices for solar abundance stars are similar for both ATK and PHOENIX models, but the Mg$_\textrm{b}$ index for the $\alpha$-enhanced model is larger for ATK. Another potential issue with modelling the Mg$_\textrm{b}$ is the presence of MgH bands in the Mg$_\textrm{b}$ index region (\citealt{Gregg94}). The strength of this molecular band is affected by 3{\small D} effects, with 1{\small D} models significantly underestimating features compared to equivalent 3{\small D} models (\citealt{Thygesen17}). The disagreement between ATK and PHOENIX predictions of C$_2$4668 indices may also be attributed to differences in the treatment of C$_2$ Swan bands (\citealt{Swan1857}; \citealt{Gonneau2016}), as is discussed in \cite{Knowles19}. The effect of geometry on Balmer lines can be also be large, as discussed in Section~\ref{sec:TestingObs}.

In summary, these comparisons show that in terms of general spectral shape and Lick line strengths, ATK and PHOENIX models predict similar differential corrections of [$\alpha$/Fe] enhancements, albeit for only two star types at solar metallicity and for a small range in [$\alpha$/Fe]. This, along with the results of \cite{Knowles19}, gives us confidence to use our models in a differential sense to correct MILES empirical stellar spectra in later sections of this work. A more important test of our models, for their application, is how well they match real star spectra. We provide further tests of our models to two different, widely-used libraries of empirical stellar spectra (see Section~\ref{sec:TestingObs}). We next describe the empirical stellar spectra that we use in the generation of a new semi-empirical library.

\section{Empirical MILES Spectra and Parameters}
\label{sec:MILESstars}
The empirical stellar spectra used in this project are from the Medium resolution Isaac Newton Library of Empirical Spectra (MILES) (\citealt{SanchezBlazquez2006}; \citealt{Falcon2011}). Whilst stars from our Galaxy do not cover the full  abundance parameter range of stars in other galaxies, they do cover a broad range in stellar parameters. MILES stars have a typical signal-to-noise of over  100\,{\AA}$^{-1}$, apart from stars which are members of stellar clusters. MILES is a stellar library for which we know attributes of $\textrm{T}_{\textrm{eff}}$, log g, [Fe/H] and [$\alpha$/Fe] for a large proportion of the whole library.

Of the 985 stars in the MILES library, \cite{Milone2011} measured the [Mg/Fe] abundances for 752 stars. We use their [Mg/Fe] measurement as a proxy for all [$\alpha$/Fe] abundances in these 752 stars for the first set of interpolations, matching MILES stars (see Section~\ref{sec:sMILESstars}). For the remaining MILES stars without [Mg/Fe] estimates, we made approximate estimates ([Mg/Fe] values of 0.0, 0.2 or 0.4) using measurements from both \cite{Milone2011} (their figure 10) and \cite{Bensby2014} (their figure 15). The \cite{Bensby2014} pattern is estimated from a study of dwarf stars in the Milky Way disk. We assigned a mean [Mg/Fe] value expected for the [Fe/H] value of the star according to the patterns found in \cite{Milone2011} and \cite{Bensby2014}. For any cluster stars that were not included in the \cite{Milone2011} work, we adopted a mean value determined for the other stars of the same cluster (see \citealt{Cenarro07} for discussion of the clusters in MILES).

The choice of which MILES stellar parameters ($\textrm{T}_{\textrm{eff}}$, log g and [Fe/H]) to use is particularly important in this work, because this will determine how well the theoretical stellar spectra and resultant semi-empirical (sMILES) spectra can represent the empirical MILES stars. These parameters will be used in interpolations within the model library to create sets of theoretical MILES stars with which to make differential corrections to empirical MILES spectra.

The two most widely-used works for MILES stellar parameters are those of \cite{Cenarro07} and Prugniel \& Sharma (\citealt{Prugniel2011}; \citealt{Sharma16}). Both sets of parameters have their benefits. In summary, the Prugniel \& Sharma parameter set has the advantage of being derived in a homogeneous fashion, from a well tested and characterised library of empirical templates, improved methodologies for lower temperature stars and good understanding of the biases involved. However, the work is limited by the use of interpolation of sometimes sparsely sampled data, particularly at the lowest temperatures where not many good star templates are available. From a bibliographic compilation, \cite{Cenarro07} produced a high-quality standard reference of atmospheric parameters for the full library of 985 MILES stars. The process involved calibrations, linked to a high-resolution reference system, and corrections of systematic differences between different sources to produce an averaged source of final atmospheric parameters from the literature, corrected to a common reference system. 

Because we plan to use the existing \cite{Vaz2015} SSP methodology in the next stage of this project, a final choice was made to use the \cite{Cenarro07} parameters, as was done previously in that work. An important reason for using \cite{Cenarro07} parameters comes from the good agreement that those parameters show with the colour-temperature-metallicity scaling of \cite{Alonso96} and \cite{Alonso99}. The SSP methodology is therefore internally consistent with the \cite{Cenarro07} parameters. In future work, there will be the possibility to use [$\alpha$/Fe] measurements currently being made for MILES stars (\citealt{GarciaPerez21}), rather than relying on the [Mg/Fe] proxy, as we are limited to currently (from \citealt{Milone2011}, [Mg/Fe] measurements).
A subsample of MILES stars were previously found not to be representative of their tagged stellar parameters. Stars were identified as problematic by matching a computed spectrum with the given stellar parameters, using the interpolator described in \cite{Vaz2010}. If the match between interpolated and observed spectrum was poor, the target star was removed from the sample or given reduced weighting in any SSP calculation that used them. These are stars with a range of issues including: low quality spectra, erroneous spectra that may have been contaminated, pointing error, spectroscopic binary, large uncertainties in stellar parameters, incorrect extinction estimates, continuum shape problems, may be a carbon star or have segments that correspond to a wrong source. These inspections are described and presented in sections 2.2 of \cite{Vaz2010} and 2.3.1 of \cite{Vaz2015}. This resulted in a final library of 925 stars for which measures of effective temperature, surface gravity and metallicity ([Fe/H]) were taken from \cite{Cenarro07} and [Mg/Fe] measures were taken from \cite{Milone2011} and estimates from \cite{Bensby2014}, as described above.

\section{Semi-Empirical MILES Library}
\label{sec:sMILESstars}
Next, we create a library of semi-empirical stellar spectra, based on application of the \textit{differential} abundance predictions of the theoretical library. This process can be split into the following steps:
\begin{enumerate}
\item{Interpolations in the theoretical MILES resolution model library to generate theoretical MILES stars. The interpolation generates spectra that exactly match MILES stars in the four atmospheric parameters of effective temperature, surface gravity, metallicity ([Fe/H]) and $\alpha$ abundance ([$\alpha$/M]=[Mg/Fe]). These are referred to as MILES theoretical base star spectra ($M_{TB}$).}
   \item{Other interpolations in the MILES model library are then made to generate theoretical MILES stars that have different abundance patterns. This interpolation matches the MILES stars in effective temperature, surface gravity and metallicity, but with different $\alpha$ abundances. These are referred to as MILES theoretical enhanced (or deficient) star spectra ($M_{T(\alpha=x)}$), where $x$ gives the [$\alpha$/Fe] abundance. For this work, $x=-0.20, 0.0, 0.20, 0.40, 0.60$.}
   \item{Differential Corrections, for each star, are then computed through :
   \begin{equation}
       \textrm{Differential Correction } (DC)=\frac{M_{T(\alpha=x)}}{M_{TB}}
       \label{DiffCorrEq}
   \end{equation}
   and are applied to empirical MILES stars to create semi-empirical MILES stars, with fluxes converted as follows, with wavelength $\lambda$:
   \begin{equation}
   \label{sMILESCalcEq}
       \textrm{sMILES}(\lambda)=\textrm{D}C(\lambda) \times \textrm{MILES}(\lambda)
   \end{equation}
 }
   \end{enumerate}
   
\begin{figure*}
\begin{center}
\includegraphics[width=160mm,angle=0]{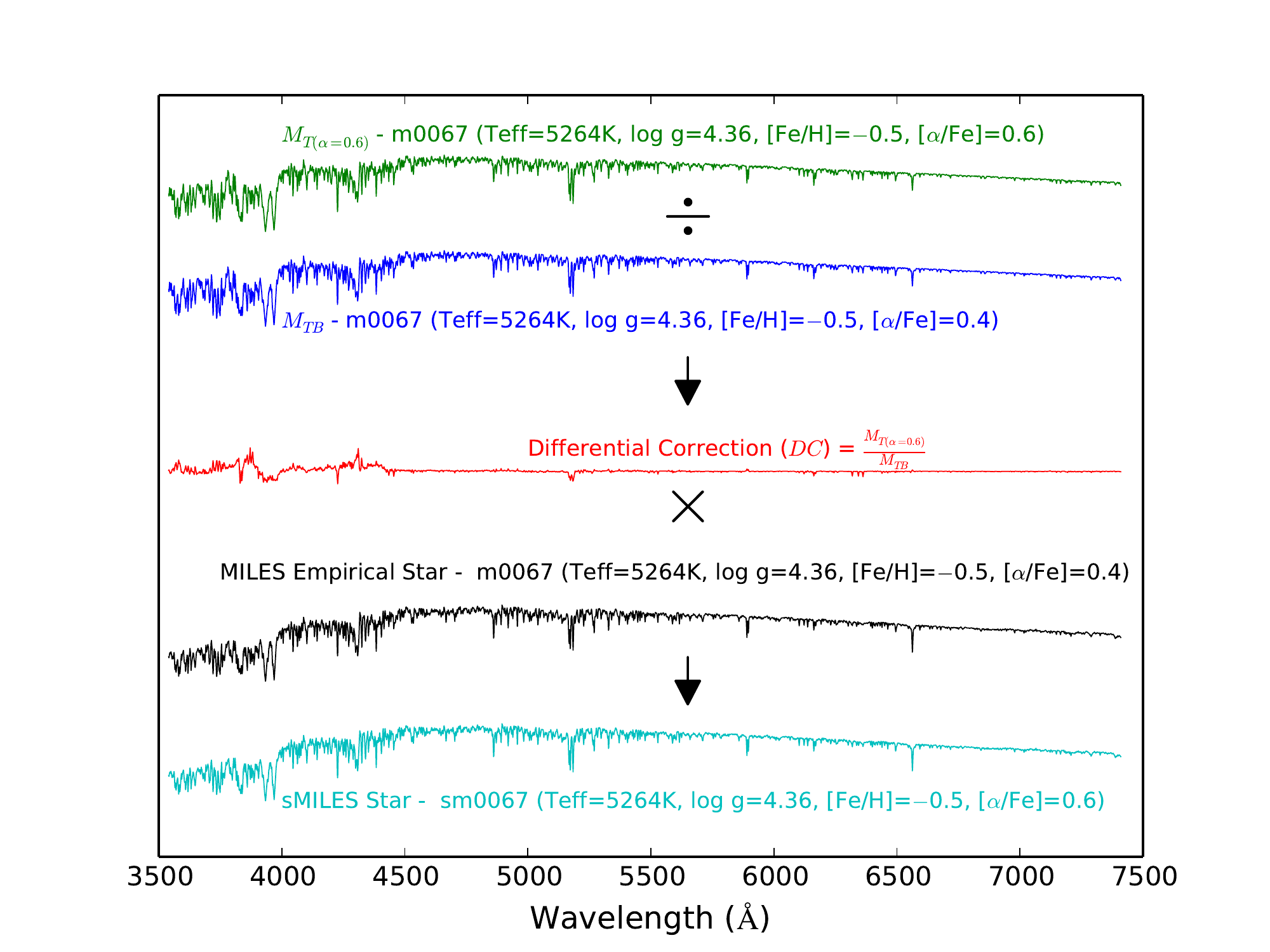}
\caption{The differential correction method followed for computing $\alpha$-enhanced and $\alpha$-deficient semi empirical (sMILES) star spectra. MILES star m0067 is shown as an example. Fully theoretical $\alpha$-enhanced ([$\alpha$/Fe]=+0.6; shown in green) and base (shown in blue) star spectra, are divided to obtain a differential correction (in red). This correction is applied to the corresponding empirical MILES star spectrum (shown in black). The result is a semi-empirical MILES star spectrum (shown in cyan) with a different [$\alpha$/Fe] ratio from the original empirical star.}
\label{m0067_V15}
\end{center}
\end{figure*}
This method produces families of semi-empirical star spectra (referred to as sMILES spectra) with the same stellar parameters (T$_\textrm{eff}$, log g and [Fe/H]) as the existing empirical MILES stars but with different abundance patterns ($\alpha$/Fe]) equal to the $M_{T(\alpha=x)}$ correction values of -0.2, 0.0, 0.2, 0.4 and 0.6. [C/Fe]=0.0 was assumed. An illustration of this process is shown in Figure~\ref{m0067_V15}, demonstrating how we apply this differential process to individual stars (rather than SSPs as in \citealt{Vaz2015}, their figure 4). We chose to perform differential corrections on stars rather than SSPs to produce a publicly available library for the community to use in their own population synthesis calculations. An alternative method for the application of differential corrections would be to produce an [$\alpha$/Fe] correction for each sampled point in a given isochrone. However, this would be dependent on the isochrone choice, and the resolution in age and metallicity of those isochrones, which may be subject to change as updates are provided. The main limitations of our chosen method here is that we are dependent on the spectral range and stellar parameter choices of the underlying empirical stellar library, which may vary, as is the case for the extended MILES library  (\citealt{Vazdekis16}) and the various determinations discussed in Section~\ref{sec:MILESstars}.

In Sections~\ref{sec:Interpolations} and \ref{sec:DiffCorr} we discuss the interpolations in the model library and the differential corrections, respectively.

\subsection{Interpolation of Theoretical Stellar Spectra}
\label{sec:Interpolations}
With MILES star parameters chosen in Section~\ref{sec:MILESstars}, the next step was to interpolate in the model library to generate theoretical spectra that match MILES stars.

To create synthetic spectra that replicate existing MILES stars, we use the interpolation mode of the software package FER\reflectbox{R}E. Designed to match spectral models to observed data in order to obtain best fitting parameters of stars, FER\reflectbox{R}E contains routines that allows interpolation within model grids. FER\reflectbox{R}E was used to interpolate in the MILES-specific theoretical library. Ratios between enhanced or deficient and base MILES star models provide the differential spectral correction (equation ~\ref{DiffCorrEq}).

The interpolation was performed using the quadratic B\'ezier function within FER\reflectbox{R}E, apart from in a few cases discussed later. A quadratic B\'ezier function is a parametric curve that is defined by three points in parameter space (e.g. in our case, the wavelength, flux density, T$_\textrm{eff}$, log g, [M/H], [$\alpha$/M] and [C/M]). The 925 star parameters were split into three groups depending on their parameters, such that they fell in the parameter range of one of the three MILES resolution and wavelength range sub-grids described in Section~\ref{sec:NewGrids}. Any stars that fell outside, or on the upper or lower grid edges were not used in the semi-empirical library. The results of these cuts meant 587, 169 and 45 stars were computed via interpolation in the 3500-6000{\small K}, 6250-8000{\small K} and 8250-10000{\small K} grid, respectively. This means that the final semi-empirical library consists of families of 801 stars with different [$\alpha$/Fe] abundances.

The first group of interpolations resulted in the MILES Theoretical Base stars, used as the denominator in the differential correction (see parameter $M_{TB}$ in equation~\ref{DiffCorrEq}). These base stars were generated by interpolating to the MILES parameters of T$_\textrm{eff}$, log g, [Fe/H] and [Mg/Fe]. Problems were found for 11 low T$_\textrm{eff}$ giant stars, for which linear interpolations were used, as described in Appendix \ref{sec:11probStars}. 

The next set of interpolations were made to produce theoretical MILES enhanced (or deficient) star spectra, used in the numerator of equation~(\ref{DiffCorrEq}). Spectra were computed with quadratic B\'ezier interpolations, in the T$_\textrm{eff}$, log g and [Fe/H] values of the existing MILES stars, but with [$\alpha$/M] values of -0.20, 0.0, 0.20, 0.40 and 0.60. This choice of [$\alpha$/M] steps reduced problems with interpolation at the grid edges, found previously. The range and sampling of [$\alpha$/Fe] abundances computed here represents an improvement over previously calculated SSP models (e.g. \citealt{Thomas05Mod}, \citealt{Conroy2012a}, \citealt{Vaz2015}) and directly relates to the range of values found in stars residing in external galaxies (e.g. figure 4 of \citealt{Sen2018}), as well as values found from unresolved stellar population studies of massive ETGs (e.g. \citealt{Conroy2014}, \citealt{McDermid15}). The 11 problem stars in the base family were computed also using linear interpolations for their $\alpha$ enhancements. The model spectra with [$\alpha$/Fe] different from those found in the local solar neighbourhood cannot easily be compared directly with real stars because they don't exist in the empirical MILES library or any other empirical libraries based on stars in the local solar neighbourhood. The result was six families of theoretical MILES stars all determined by interpolation of the model grids - one with all the existing MILES parameters and five with the same fundamental parameters but different $[\alpha$/Fe] abundances on a regular grid, at MILES resolution and wavelength range.

\subsection{Differential Corrections}
\label{sec:DiffCorr}
\begin{figure*}
\centering
 \includegraphics[width=\linewidth, angle=0]{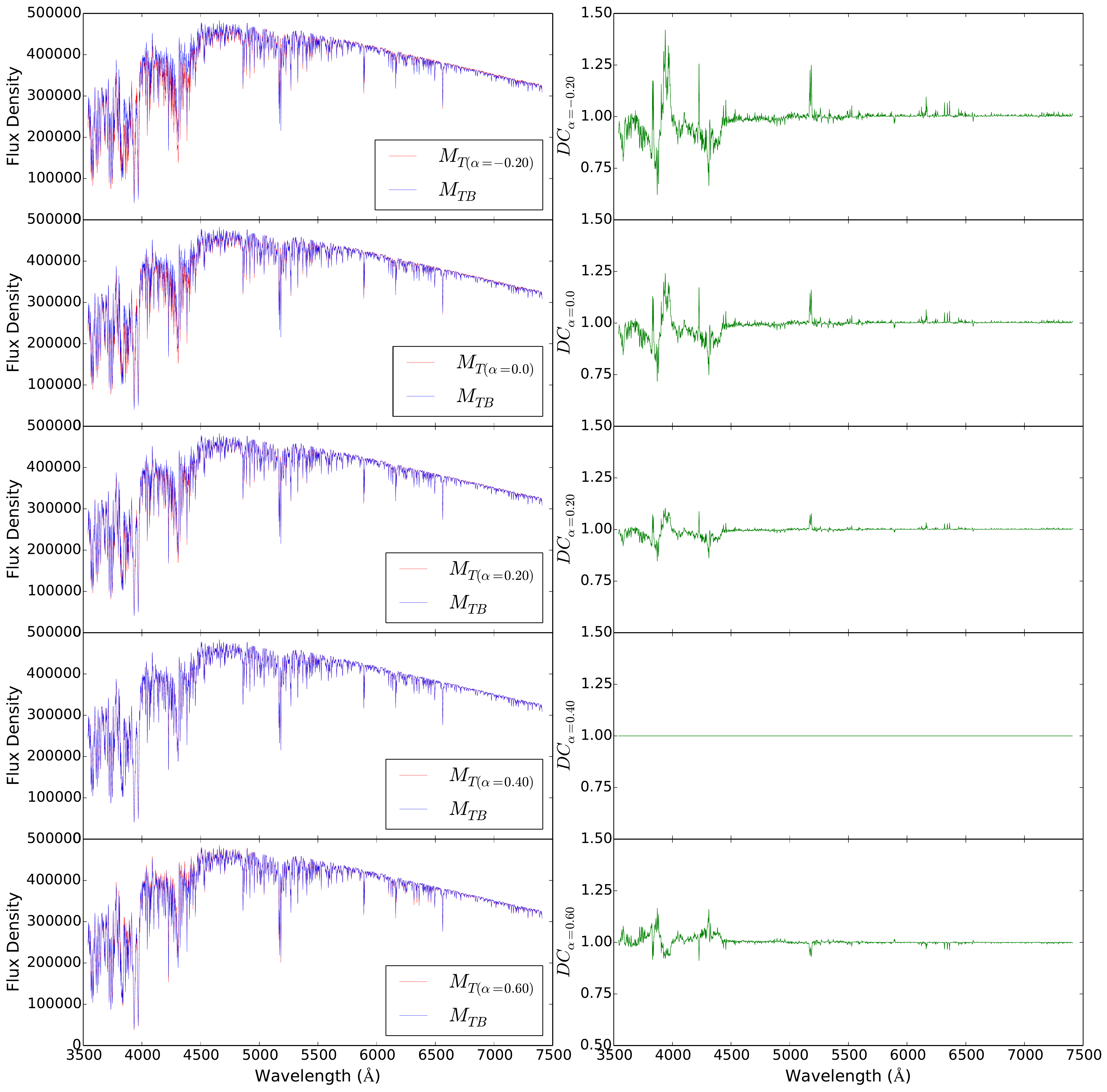}
 \caption{Example differential corrections, which are applicable for MILES star m0067 (=HD010700: T$_\textrm{eff}$=5264{\small K}, log g=4.36, [Fe/H]=-0.50, [Mg/Fe]=0.40). The left panel compares the resulting spectra of theoretical enhanced (or deficient) ($M_{T(\alpha=x)}$) and theoretical base $M_{TB}$ stars. In these plots ($\alpha$=x) is short for ([$\alpha$/Fe]=x). Flux Density is in units of erg/s/$\textrm{cm}^2$/\AA. The right panel shows the resulting differential correction ($\textrm{DC}_{(\alpha=x)}$), derived from equation~(\ref{DiffCorrEq}), for each of the output [$\alpha$/Fe] abundances. Note that for this star, the differential correction for [$\alpha$/Fe]=0.40 is 1, because the empirical MILES star is already at [Mg/Fe]=0.40.}
 \label{ATK_DiffCorr_m0067}
\end{figure*}
\begin{figure*}
\centering
 \includegraphics[width=\linewidth, angle=0]{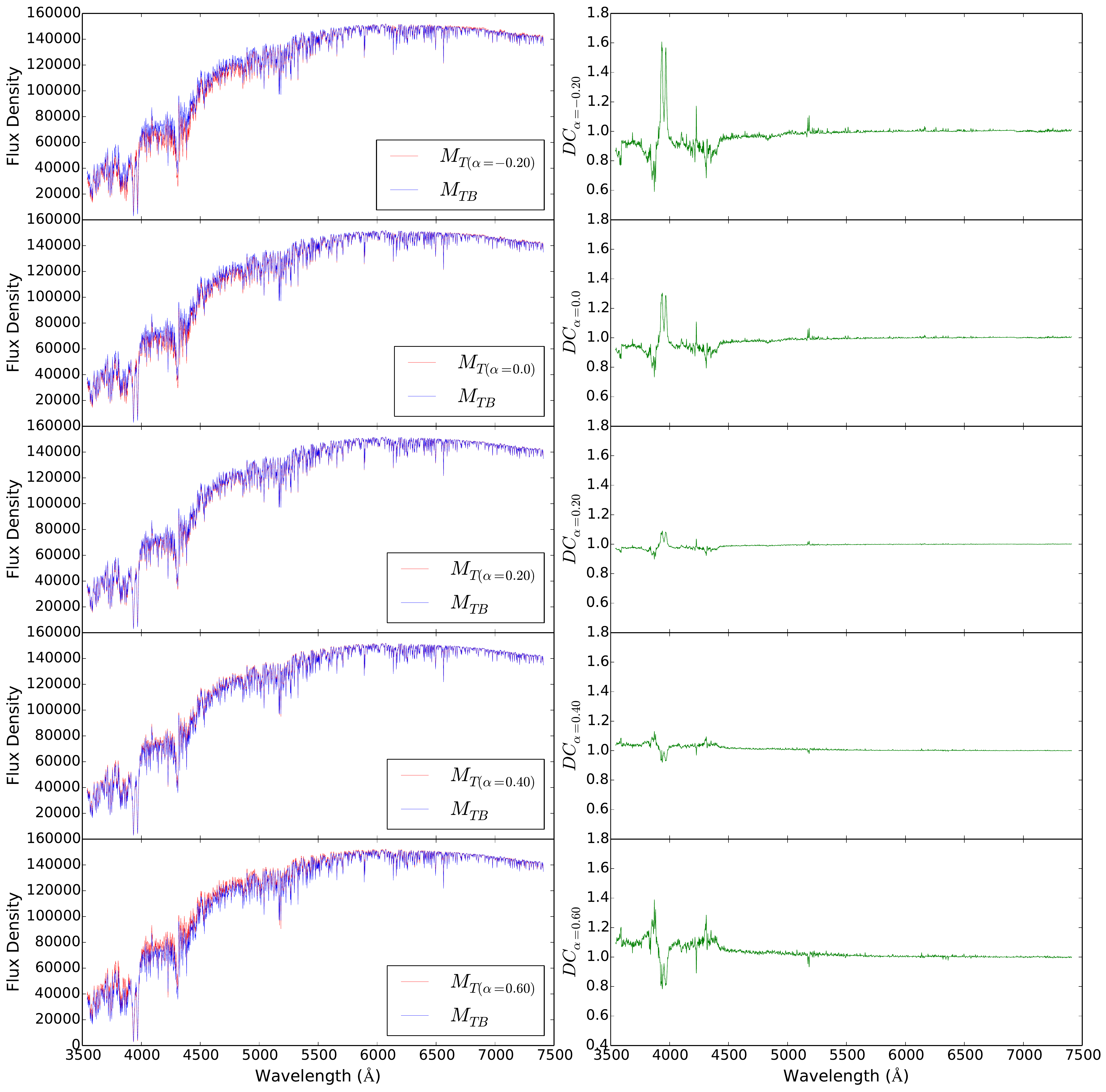}
 \caption{Example of Differential corrections, which are applicable for MILES star m0923 in the globular cluster M3 (=M3 IV 25: T$_\textrm{eff}$=4367{\small K}, log g=1.27, [Fe/H]=-1.34, [Mg/Fe]=0.30). This empirical MILES star has an [Mg/Fe] value of 0.3. The left panel compares the resulting spectra of the theoretical enhanced (or deficient) ($M_{T(\alpha=x)}$) and theoretical base $M_{TB}$ stars. In these plots ($\alpha$=x) is short for ([$\alpha$/Fe]=x). Flux Density is in units of erg/s/$\textrm{cm}^2$/\AA. The right panel shows the resulting differential correction ($\textrm{DC}_{(\alpha=x)}$), derived from equation~(\ref{DiffCorrEq}), for each of the output [$\alpha$/Fe] abundances.}
 \label{ATK_DiffCorr_m0923}
\end{figure*}

 Python routines performed the division of flux of the enhanced (or deficient) over base spectra, described in equation~(\ref{DiffCorrEq}) for each wavelength. In equation~(\ref{DiffCorrEq}) $\alpha$ indicates the [$\alpha$/Fe] abundance of the sMILES star that will be produced if the differential correction is applied to the empirical MILES star. Two example sequences of differential corrections are shown for MILES stars m0067 (T$_\textrm{eff}$=5264{\small K}, log g=4.36, [Fe/H]=$-$0.50, [Mg/Fe]=0.40) and m0923 ((T$_\textrm{eff}$=4367{\small K}, log g=1.27, [Fe/H]=$-$1.34, [Mg/Fe]=0.30) in Figures~\ref{ATK_DiffCorr_m0067} and ~\ref{ATK_DiffCorr_m0923}, respectively. As shown, the differential correction is smallest for abundance patterns closest to the measured value of the empirical star. The largest differential corrections are found for wavelengths below $\sim4500\,${\AA}. This is likely due to the many strong metal line and molecular features that increasingly accumulate below $\sim4500\,${\AA}, such as the G-band, Ca H\&K, CH and CN contributions. \cite{Coelho05} (their figure 16) shows the increasing contributions from atomic lines and certain molecular bands at these shorter wavelengths. Figure 2 of our supplementary data demonstrates the effect of metallicity at these shorter wavelengths in our own models. The effects of [$\alpha$/Fe] enhancements shown here are in agreement with previous works (e.g. \cite{Cassisi2004}, their figure 2). Another noticeable feature in the corrections is also present around the Mg$_\textrm{b}$ indices, which again increases as the $[\alpha$/Fe] abundances differ from the measured abundance of the empirical star.

The appropriate differential correction was then applied to the corresponding empirical star spectrum value via equation~(\ref{sMILESCalcEq}). The result was 801 spectra in each of the five [$\alpha$/Fe] bins, with a wavelength coverage of $3540.5 - 7409.6\,${\AA} in bins of $0.9\,${\AA}. 

\begin{figure*}
\centering
 \includegraphics[width=140mm, angle=0]{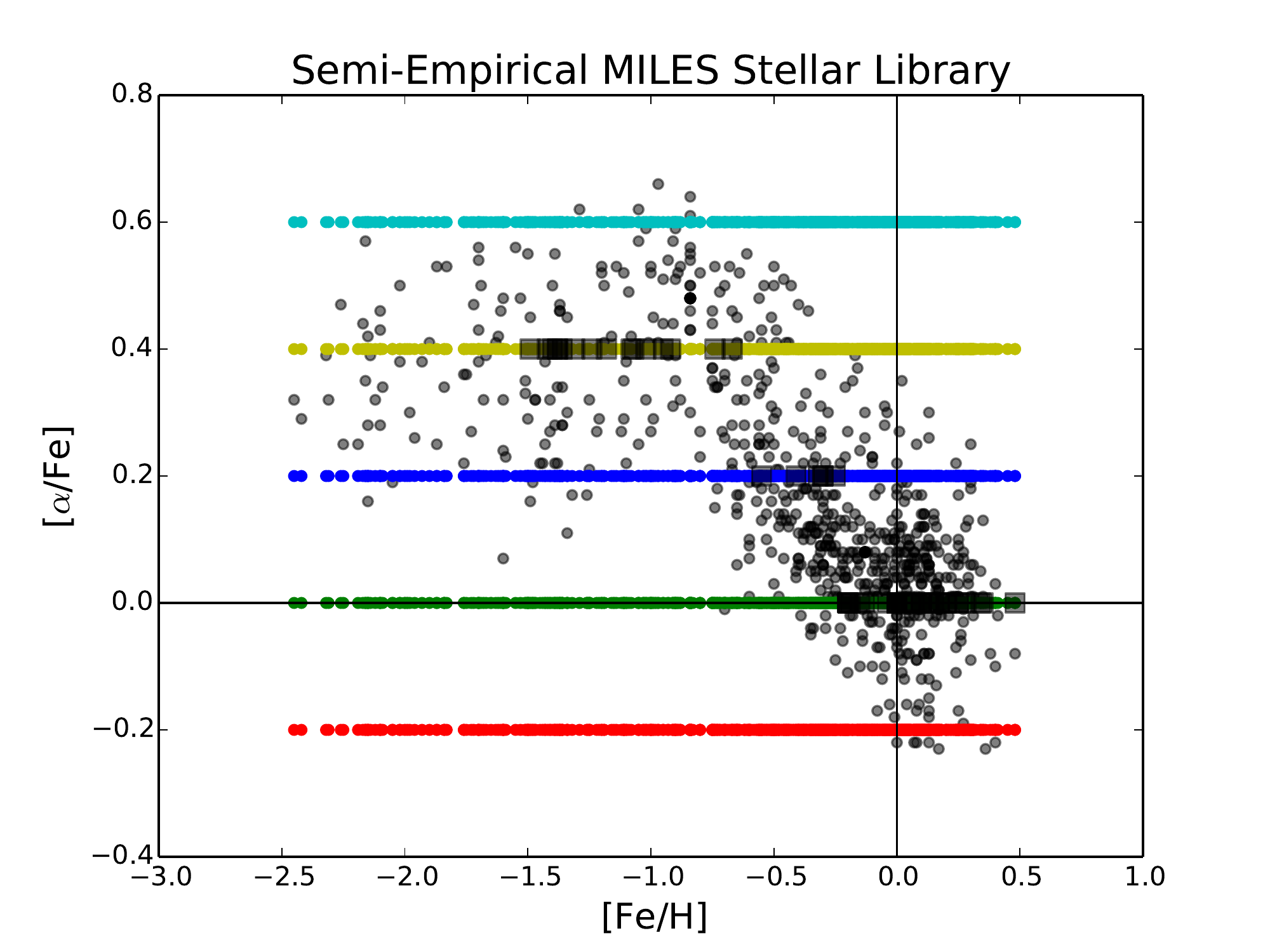}
\caption[Final sMILES stellar library produced in this work in $\alpha$ abundance and metallicity, in comparison to the empirical MILES library] {Final semi-empirical MILES (sMILES) stellar library coverage in the [$\alpha$/Fe] vs [Fe/H] plane. The coloured points that lie in horizontal lines represent the families of 801 sMILES stars and black points represent the corresponding 801 empirical MILES stars, with black squares representing those stars with [Mg/Fe] values estimated from a Milky Way relation derived in \cite{Bensby2014} and black circles representing the 752 MILES stars for which \cite{Milone2011} provides [Mg/Fe] values. T$_\textrm{eff}$, log g and [Fe/H] values were taken from \cite{Cenarro07}.}
\label{sMILES_Library}
\end{figure*}

To summarise the sMILES library, we plot the locations of sMILES stars in the [$\alpha$/Fe] vs [Fe/H] plane in Figure~\ref{sMILES_Library} to show the final coverage in these parameters and 801 empirical MILES stars, which show the well known distribution of abundances, for stars in the local solar neighbourhood. Each horizontal coloured line represents a family of 801 sMILES stars at a given [$\alpha$/Fe]. Similar figures are provided in the supplementary material to show the coverage of sMILES stars in the [$\alpha$/Fe] vs T$_\textrm{eff}$ and log g planes. Next we test our new theoretical spectra and differential corrections against existing observed spectra from different empirical libraries. 

\section{Testing Model Spectra and Differential Corrections against Real Stars}
\label{sec:TestingObs}
As indicated in Section~\ref{sec:sMILESstars}, it is difficult to test the full range of our sMILES library, because not all such parameter combinations can be found in nearby stars (see Figure~\ref{sMILES_Library}). The fact that a wider range of abundance parameter combinations do appear to exist elsewhere in the Universe (e.g. in dwarf spheroidals and giant ellipticals) is the reason why we wished to generate these sMILES spectra. However, we can do some limited tests. We first compare our theoretical grid to empirical MILES stellar spectra. Then we compare our theoretical spectra and differential corrections to spectra selected from the empirical MaStar stellar library.

\subsection{MILES Comparisons}

\begin{figure*}
\centering
\includegraphics[width=147mm, angle=0]{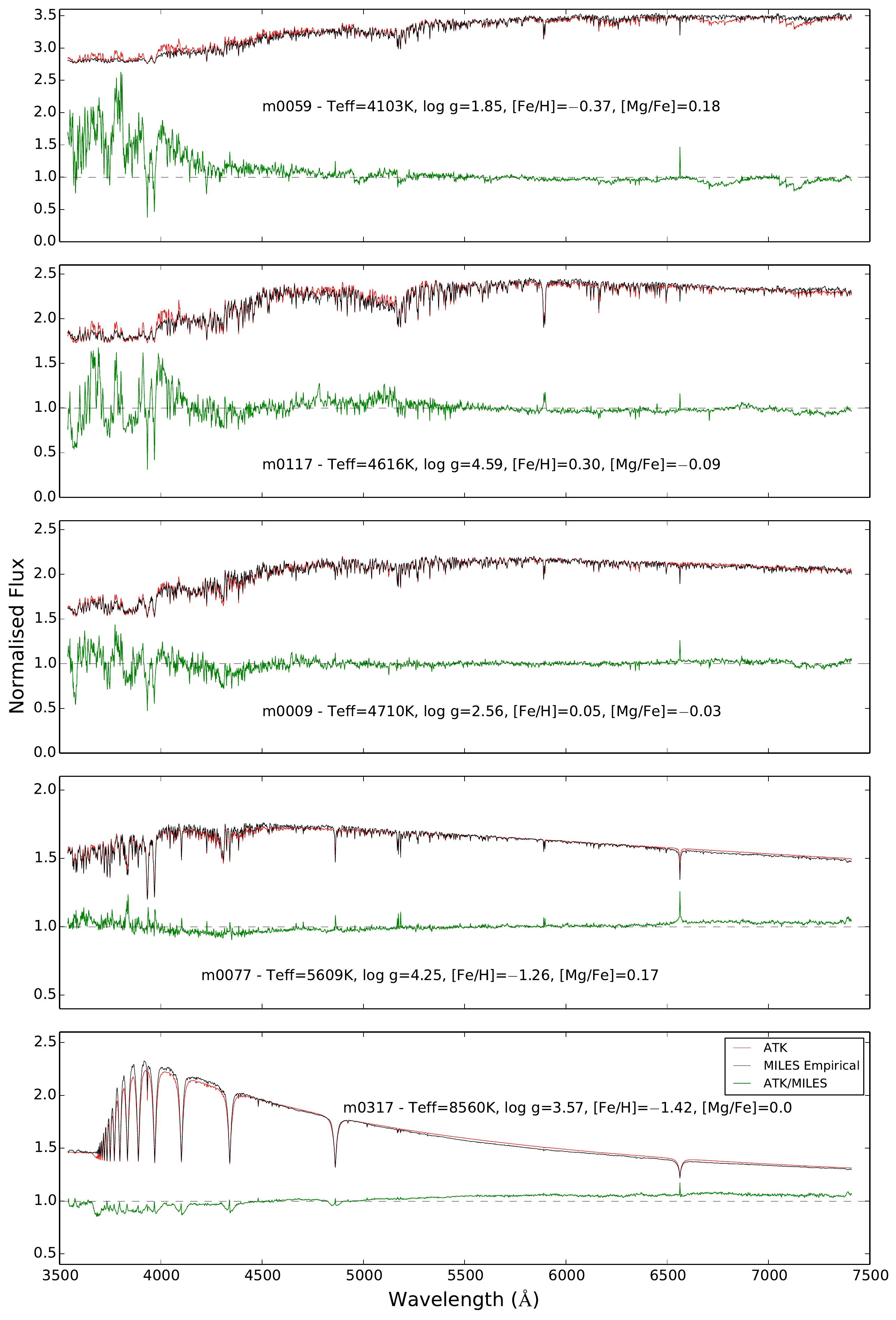}
\caption{Comparison of ATK model and empirical MILES stars for m0009, m0059, m0077, m00117, m0317. MILES star parameters from \protect\cite{Cenarro07} and \protect\cite{Milone2011} are given in each panel. Spectra are degraded to $2.5\,${\AA}, sampled at $0.9\,${\AA} and normalised to unity area. ATK (red lines) and MILES (black lines) spectra are scaled up by a factor of 2000 and shifted onto the plots. Ratios between ATK models and MILES stars are given in the lower panel of each plot (green lines) with no scaling or shifting applied. The vertical axes on the plots are to scale for the ratios between ATK models and MILES stars. The 1:1 agreement between ATK models and MILES empirical star is represented as a dashed-horizontal line.}
\label{ATK_vs_MILES_5stars}
\end{figure*}

\begin{figure*}
\centering
\includegraphics[width=150mm, angle=0]{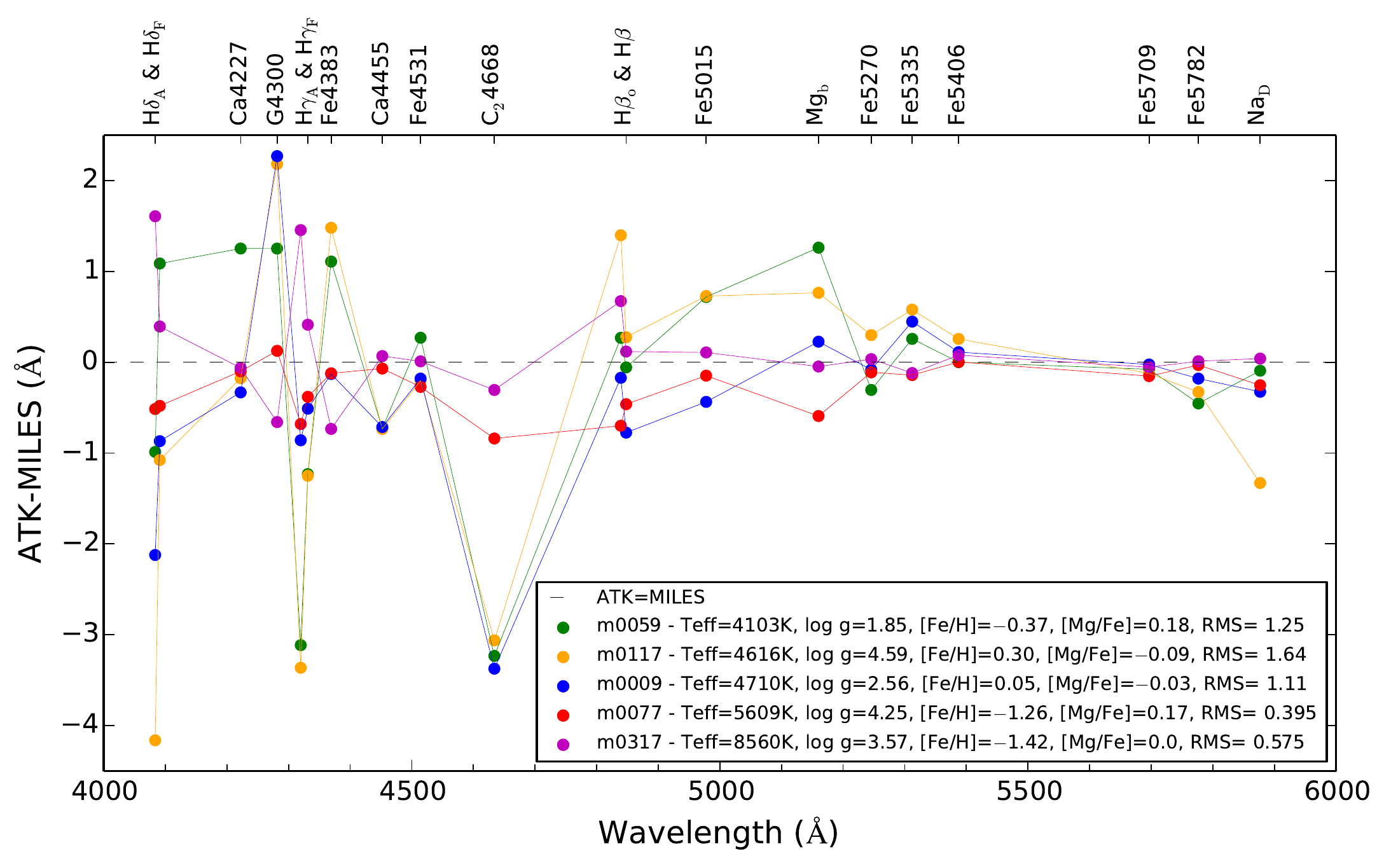}
\caption{Comparison of Lick indices predicted by our interpolated MILES models (labeled ATK) to empirical MILES stars. This is for indices that are measured in {\AA}, including H$\beta_{\textrm{o}}$ from \protect\cite{Cervantes09} and for five examples of stars present in the MILES library. Lick indices are labelled for illustration and the 1:1 agreement between model and observation is plotted as a dashed horizontal line. Disagreements between models and observations are generally larger in the blue. RMS scatter (in {\AA}) about the 1:1 agreement line is given for each star.}
\label{ATK_vs_MILES_Lick}
\end{figure*}

 Although the MILES stars will reflect the Milky Way abundance pattern, checks can still be made to test the model grid in various parts of parameter space. To test models directly to MILES stars, we use the theoretical MILES base stars, generated through quadratic interpolations within the model grids, as described in Section~\ref{sec:sMILESstars}.

In Figure~\ref{ATK_vs_MILES_5stars}, we show comparisons of these models to MILES stars for various star types, specifically with varying metallicities and [Mg/Fe] values. The cool stars show increasingly larger differences below about 4200{\AA} (e.g. m0059, m0117). The sharp cores of hydrogen alpha lines are not well reproduced in any of the theoretical spectra. Balmer lines in general are poorly fit for the higher temperature stars (e.g. m0317). Ca H\&K lines are stronger in the theoretical models for cool stars than in the MILES stars (e.g. m0059, m0117). The coolest star model (m0059) also shows a mismatch in the red, with molecular features stronger in the theoretical model compared with the MILES star. These results are in agreement with the findings of \cite{Knowles19}, with differences between observations and models identified for cool stars. In Figure~\ref{ATK_vs_MILES_Lick} we also show the differences between predicted Lick indices for our interpolated MILES models and the equivalent empirical MILES stars.

In general, the agreement between models and MILES stars is worst at the bluer wavelengths of the MILES range, with reasonable agreements found above $\sim4500\,${\AA}. This is as expected from previous direct comparisons, which have also shown wavelength-dependent disagreements between theoretical models and observed spectra  (e.g. \citealt{Martins07}; \citealt{Bertone08}; \citealt{Coelho2014}; \citealt{Villaume17}; \citealt{Allende18}); \citealt{Knowles19}). The models tested here are generated using versions of ATLAS therefore, spherical geometry and non-LTE effects have been ignored. These assumptions may explain the lack of agreement between models and observations, particularly for cool stars. The absolute effect of spherical geometry, in the form of convection, on Balmer lines can be large, resulting in differences between 3{\small{D}} LTE and 1{\small{D}} LTE temperature estimates of late-type stars of up to $\approx$200{\small{K}} (Table 4 of \citealt{Amarsi2018}). Balmer lines modelled under LTE conditions are known to match the line wings, but cannot reproduce the core of the lines (e.g. figures 5 and 6 in \citealt{Amarsi2018} and section 4.2 in \citealt{Martins07}). However, the effect of non-LTE in the cooler temperature regimes tested here are smaller than the 3{\small D} effects, particularly for higher order Balmer features (Table 4 of \citealt{Amarsi2018}). Generally, non-LTE effects become more important in the very lowest and highest temperature stars, in addition to very metal poor stars or those with low surface gravity (e.g. \citealt{Hauschildt99a}; \citealt{Martins2005}; \citealt{Hansen2013} and references therein). The disagreement between model and observed hydrogen line indices (see Figure~\ref{ATK_vs_MILES_Lick}) may also be partly explained by the presence of chromospheres, which can reduce the absorption or even producing emissions in the cores of Balmer lines (e.g. \citealt{Leenaarts12}), and limitations in the atomic data in the region. The Balmer lines in cool stars can be weak and the region could be affected by poorly-known, uncalibrated metal lines. We highlight again here that in the generation of sMILES stars (Section~\ref{sec:sMILESstars}), we use the models in a differential sense only. In the application of predictions of these models, we have shown that using the models' differential predictions of abundance pattern effects produces a better agreement with observations than using the absolute predictions, particularly at bluer wavelengths (\citealt{Knowles19}, their figure 11). The differential predictions of some hydrogen features are scattered by a factor of $\sim$2 less than the absolute predictions and a large reduction in scatter between the two approaches is also seen in G4300 and C$_{2}$4668 indices. Another potential source of disagreement between models and observations here is any abundances differences other than [$\alpha$/Fe], such as C and N, which might affect the empirical stars but are not changed from scaled-solar in the interpolated models. C and N have quite a large effect on the spectra, particularly in the blue (see Response Tables of \citealt{Knowles19}). Future improvements would involve modelling more individual elements in the theoretical models and more accurate measurements of their abundances in empirical stellar spectral libraries.
Next, we test our theoretical models to a more recent set of observations.

\subsection{MaStar Comparisons}
A recent large survey of stars in our Solar neighbourhood is that of the MaStar empirical stellar spectral library (\citealt{Yan2019}). These spectra, covering $3622-10354\,${\AA}, were observed using the BOSS spectrograph on the 2.5m SLOAN telescope at Apache Point Observatory. They obtained good quality spectra for 3321 stars, with spectral sampling of $\Delta\log(\lambda\,$(\AA)$)=1\times10^{-4}$, corrected to rest-frame vacuum wavelengths and flux calibrated, but uncorrected for foreground Galactic extinction. The spectral resolution varies with wavelength, and between observations, as shown in \cite{Yan2019} their figure 10. Typically, the spectral resolution of the MaStar observations is $\sim3\,${\AA} (FWHM), at wavelengths up to $\sim6000\,${\AA}, and increases non-linearly to $\sim5\,${\AA} (FWHM) at the reddest wavelengths. There are 1589 MaStars with [$\alpha$/Fe] measurements, in addition to [Fe/H], $\textrm{T}_{\textrm{eff}}$ and log g measurements, available from their input stellar parameter catalogues from APOGEE, SEGUE and LAMOST surveys (see \citealt{Yan2019} for details).

With these stellar parameter measurements for 1589 stars, this makes the MaStar spectral catalogue a potentially useful resource for comparing with our theoretical star spectra, independently of the MILES stellar library. Therefore we compare MaStar spectra, extracted from the MaStar good spectral catalogue\footnote{\url{https://data.sdss.org/sas/dr16/manga/spectro/mastar/v2\_4\_3/v1\_0\_2}}, with our new theoretical star spectra. 

\begin{table}
\caption{Selection parameters showing values for four theoretical stars and ranges about those values (last row) for selection of observed stars from the MaStar good spectral catalogue, with good quality flag MJDQUAL=0. Cool giant (CG) and cool dwarf (CD) stars are listed. CG\_e and CD\_e are more enhanced cool giant and cool dwarf stars.}
\begin{tabular}{llllll}
\hline
\multicolumn{4}{l}{VALUES AND (RANGES)} & \multicolumn{2}{l}{OBSERVED STARS}\\
T$_\textrm{eff}$ & [Fe/H] & log g & [$\alpha$/Fe] & Number and & Number\\
({\small K})  & (dex)     & (dex) & 
(dex) & Type of Stars & of Spectra\\
\hline
%\multicolumn{4}{l}{Cool Giants (CG)} &&\\
4750 & -0.4 & 2.5 & +0.05 & 4 CG & 12\\
     &      &     & +0.20 & 6 CG\_e & 12\\
%\multicolumn{4}{l}{Cool Dwarfs (CD)} &&\\
     &      & 4.5 & +0.05 & 4 CD & 15\\
     &      &     & +0.20 & 7 CD\_e & 19\\
%\hline
%\multicolumn{4}{l}{RANGES} &&\\
%$\Delta$(T$_\textrm{eff}$) & $\Delta$[Fe/H] & $\Delta$(Log g) & $\Delta$[$\alpha$/Fe] &&\\ 
%({\small K}) & (dex) & (dex) & (dex) &&\\
%\hline
($\pm$100) & ($\pm$0.1) & ($\pm$0.2) & ($\pm$0.06) &&\\
\end{tabular}
\label{Example MaStars}
\end{table}

To investigate effects of individual parameters, we selected groups of MaStars that lie within small errors from specific theoretical stars. Errors on abundance parameters ([Fe/H] and [$\alpha$/Fe]) are large for any one star, typically $\pm\sim$0.05 to $\pm$0.1 dex (e.g. for SEGUE spectra in \citealt{Lee2011}), plus uncertain systematic errors. By selecting groups of similar stars we aim to reduce the uncertainty in their average abundances. The parameters chosen were guided by the wish to test differential effects of [$\alpha$/Fe]. This constraint limits the parameter space from which we can select groups of stars in the our Solar neighbourhood because the range of [$\alpha$/Fe] is small at any given value of [Fe/H]. In \cite{Yan2019}, their figure 13, we see that the best place to look for groups of similar stars is at slightly sub-solar metallicity of [Fe/H]$\sim$-0.4, where there is a group of stars at [$\alpha$/Fe]$\sim$+0.05 and another group at [$\alpha$/Fe]$\sim$+0.2 that we hereafter refer to as the enhanced group. We selected cool MaStars ($\textrm{T}_{\textrm{eff}}\sim$4750{\small K}) with values and ranges detailed in Table~\ref{Example MaStars}, around these abundances, and sampled two values of log g.

Four theoretical spectra with parameters given in Table~\ref{Example MaStars} were created, using FER\reflectbox{R}E interpolation of the model grids as elsewhere in this paper. Although the MaStar spectra are flux calibrated, they show variations from multiple observations of the same star that need to be removed in order to make the comparisons with our theoretical spectra. We chose a weighting for the continuum fit that would de-emphasise the absorption features. Therefore, the spectra were processed as follows, using Python code and IRAF routines:

\begin{itemize}
\item{Flattened by division of a fourth order Legendre polynomial fit, weighted by flux squared (MaStar and theoretical spectra).}
\item{Smoothed to $3\,${\AA} FWHM resolution (theoretical spectra), to approximately match MaStar resolutions.}
\item{Converted to air wavelengths (MaStar spectra), so that all spectra are on the same wavelength scale.}
\item{Binned to $1.0\,${\AA} linear bins (MaStar and theoretical spectra).}
\end{itemize}

\begin{figure*}
\centering
\vspace{-0.85cm}
\subfloat{\includegraphics[width=96mm, angle=0]{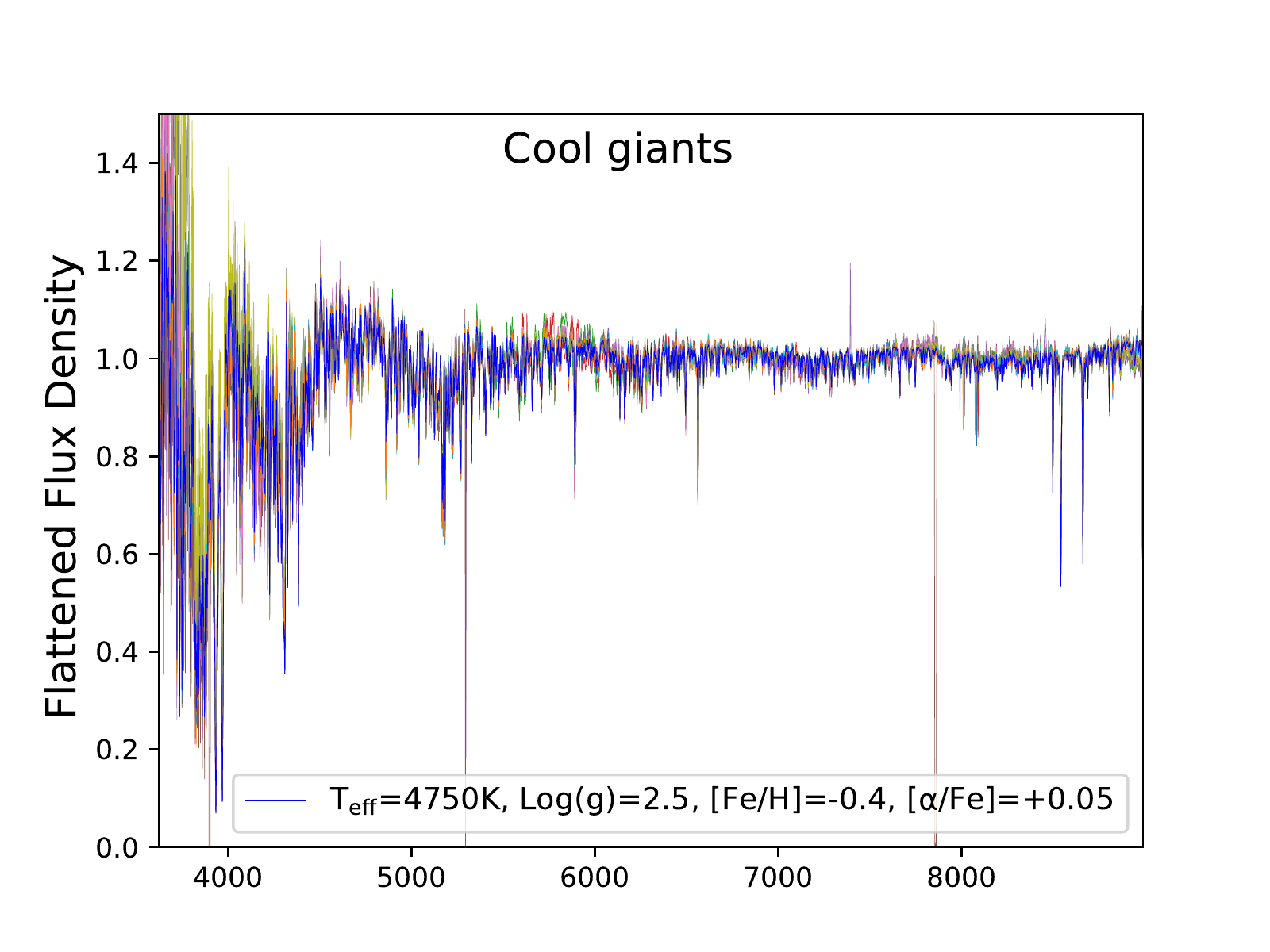}}
\subfloat{\hspace{-0.9cm}\includegraphics[width=96mm, angle=0]{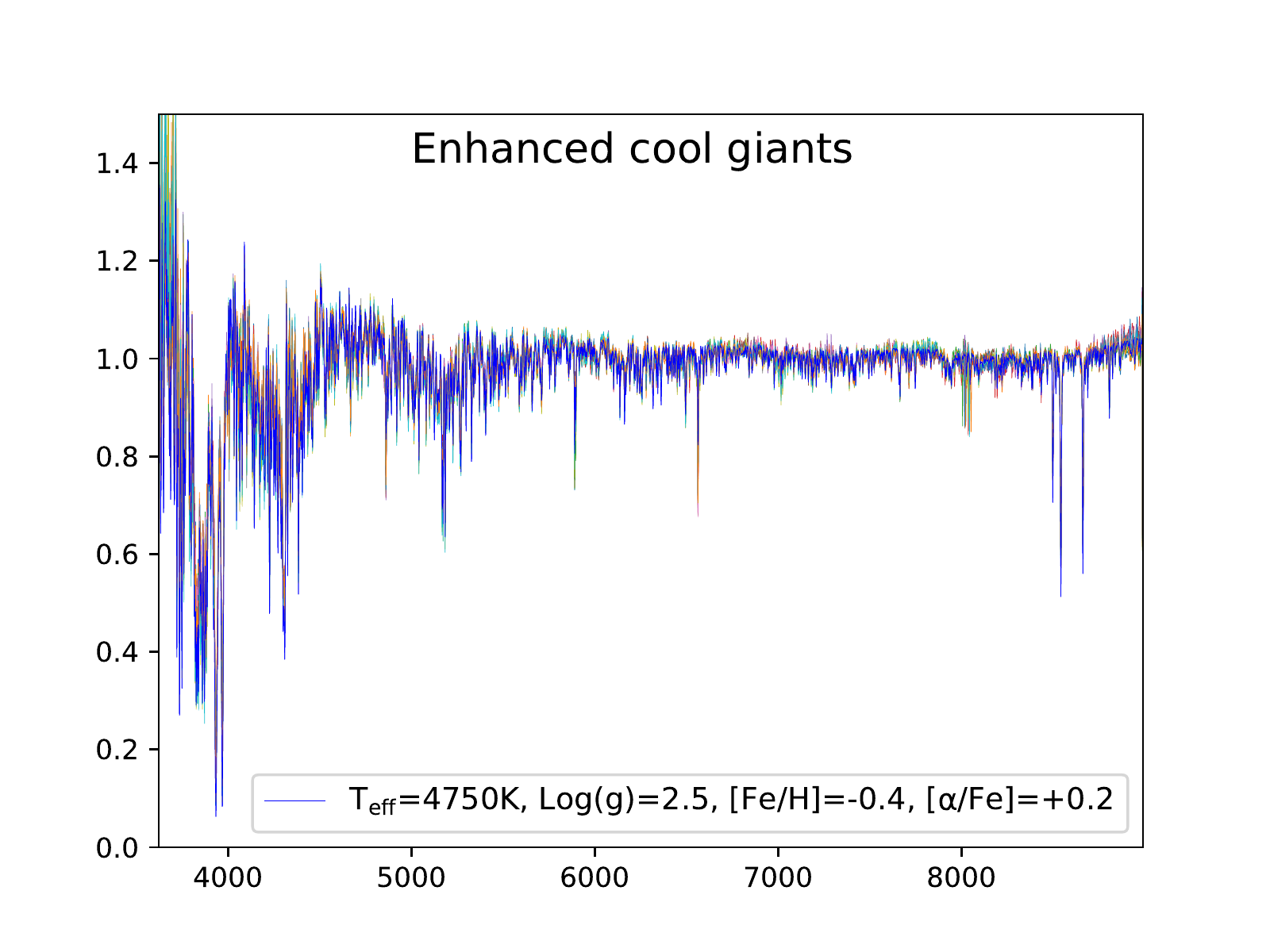}}
\vspace{-0.85cm}
\subfloat{\includegraphics[width=96mm, angle=0]{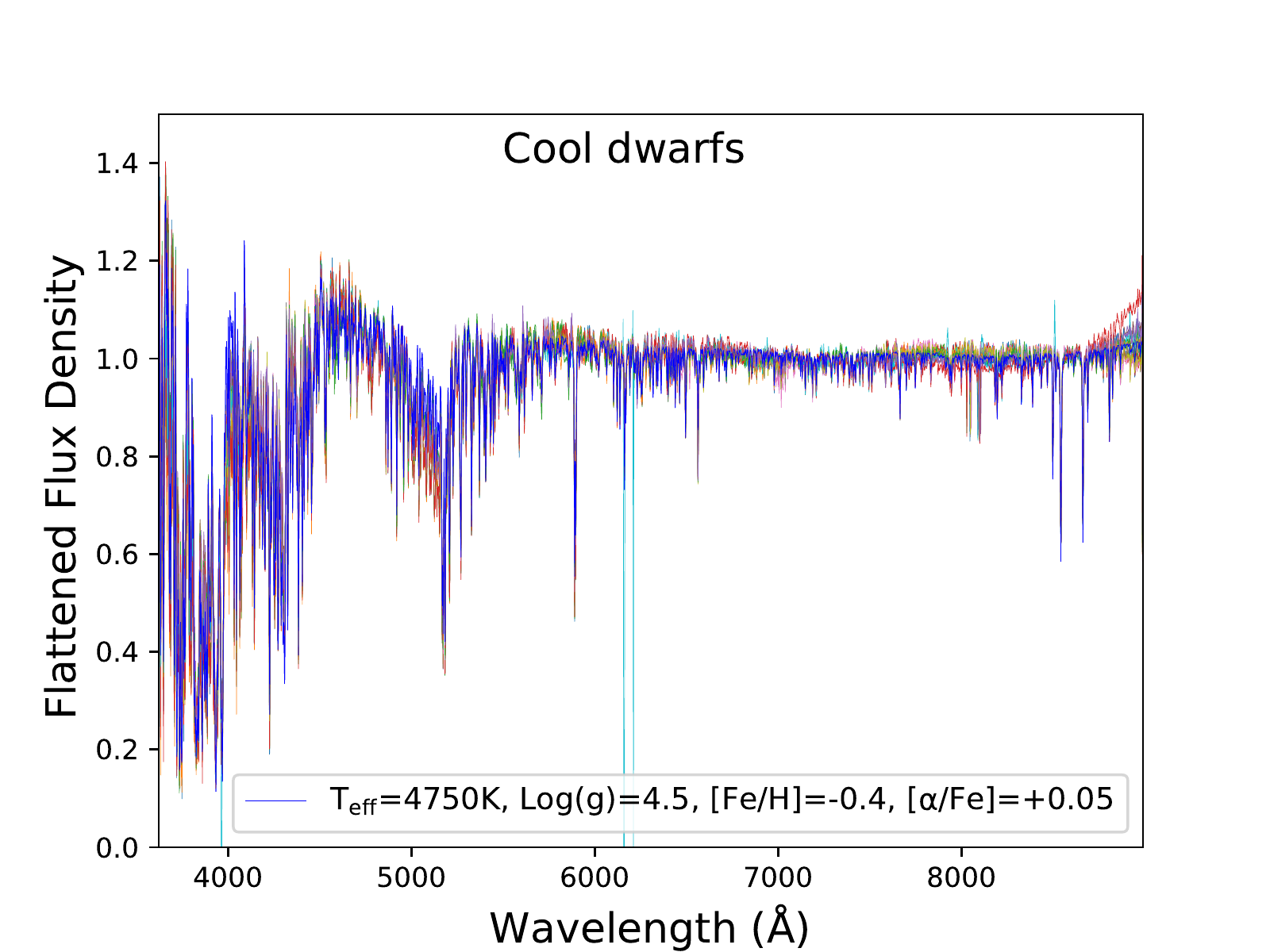}}
\subfloat{\hspace{-0.9cm}\includegraphics[width=96mm, angle=0]{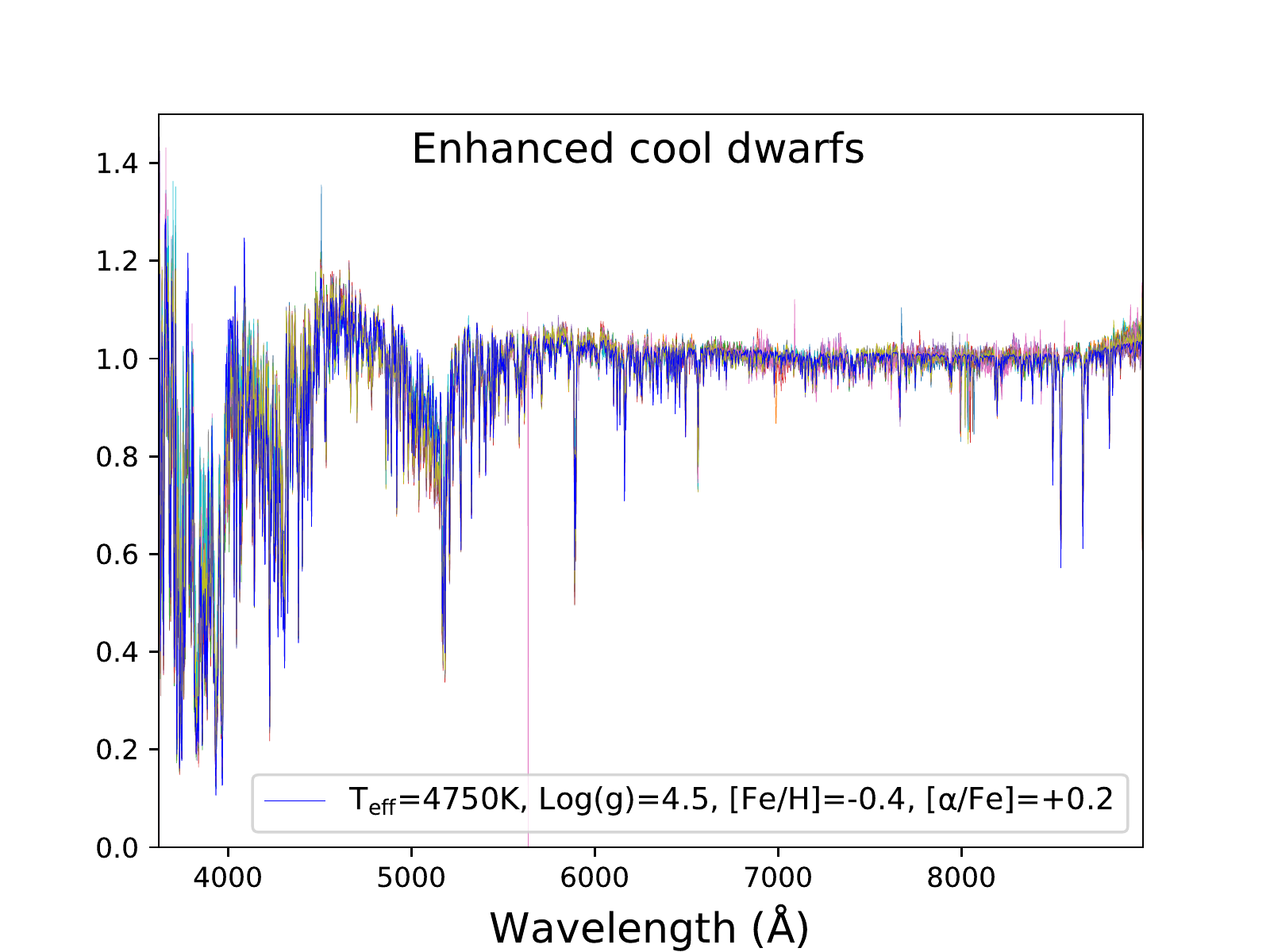}}

\caption{Flattened spectra of our theoretical stars (blue line) overlaid on those of empirical MaStar spectra (multiple coloured thin lines), for the four spectral types listed in Table~\ref{Example MaStars}. The upper two panels show cool giant stars at [$\alpha$/Fe]=+0.05 (left panel) and [$\alpha$/Fe]=+0.20 (right panel). The lower two panels show cool dwarf stars at [$\alpha$/Fe]=+0.05 (left panel) and [$\alpha$/Fe]=+0.20 (right panel).}
\label{ATK_vs_MaStar_Spec}
\end{figure*}

\begin{figure*}
\centering
\includegraphics[width=125mm, angle=0]{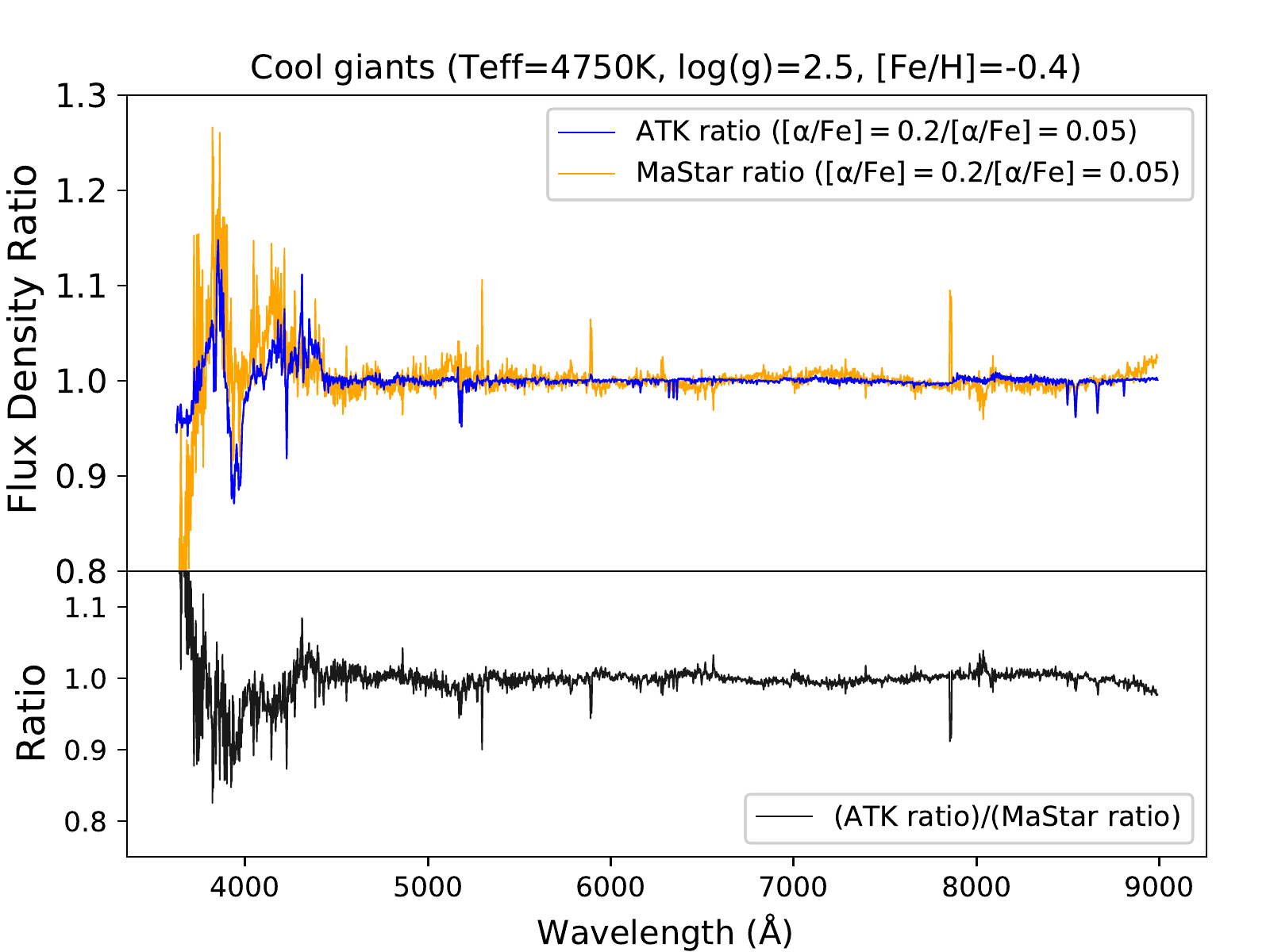}
\includegraphics[width=125mm, angle=0]{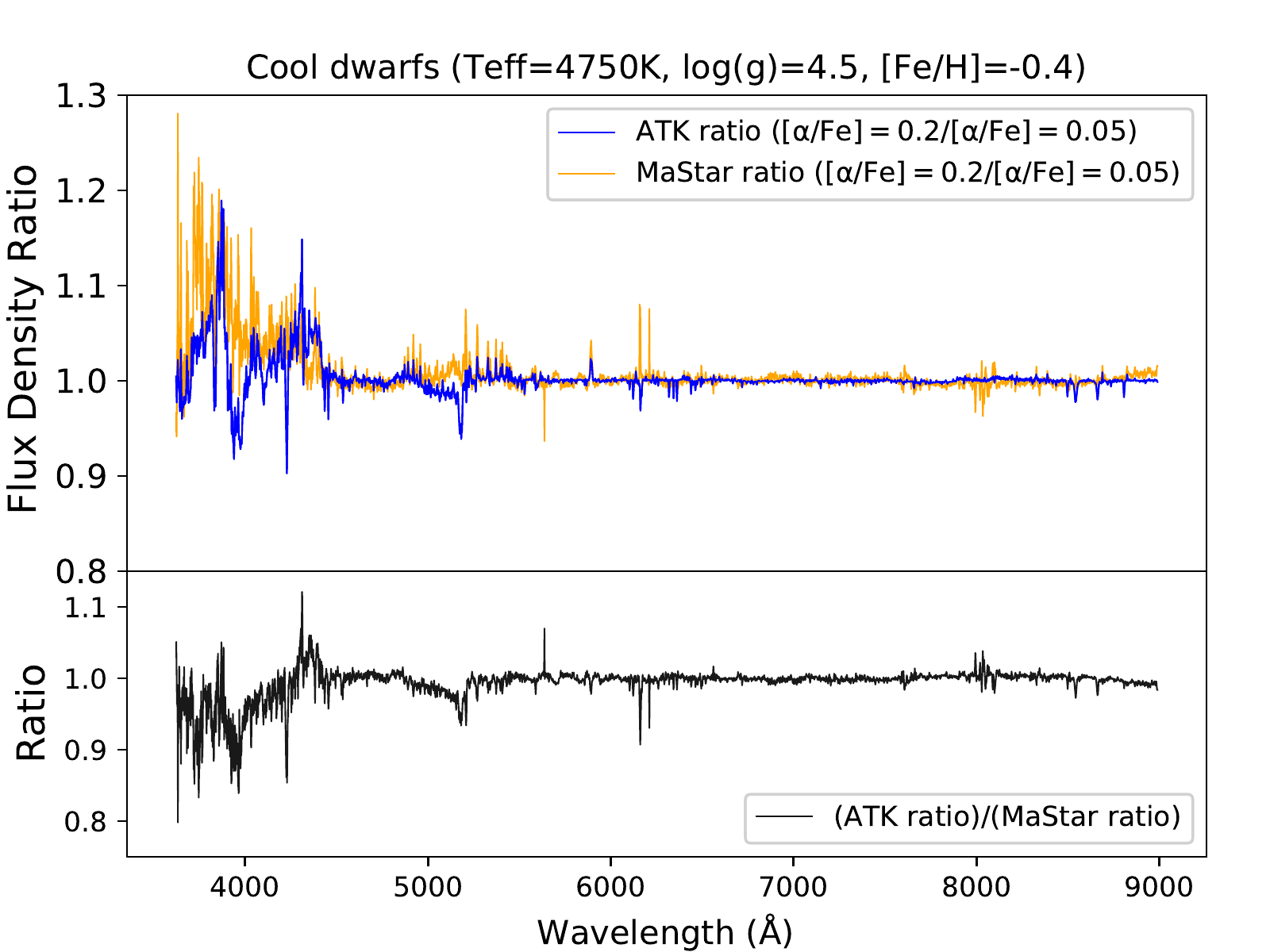}
\caption{Top plot: Differential enhancements in cool giant stars, shown by flux density ratios of enhanced to less-enhanced flattened star spectra for our theoretical stars (dark blue lines, labelled ATK ratio) and for averaged cool giant MaStars (orange lines, labelled MaStar ratio). The lower panel (black line) shows the division of these ratios, highlighting residual mismatches between theory and observations in their differential changes due to [$\alpha$/Fe] enhancements. Bottom plot: The same, but for cool dwarf stars.}
\label{ATK_vs_MaStar_DiffCorr}
\end{figure*}

In Figure~\ref{ATK_vs_MaStar_Spec} we show the resultant theoretical star spectrum (dark blue line), overlaying the corresponding MaStar spectra (multiple coloured, thin lines), for each of the four star types listed in Table~\ref{Example MaStars}. Figure~\ref{ATK_vs_MaStar_Spec} shows that the spectral structures agree well, after flattening, and the difference between giant (upper row) and dwarf (lower row) stars is clear, for both the theoretical and observed star spectra. These trends of deepening features around the magnesium band and sodium doublet lines in cool dwarfs are the same as seen in \cite{Knowles19_Thesis} (figures 3.13 and 3.14), whilst the near-IR calcium triplet lines go in the opposite sense, getting weaker at higher surface gravity, as in \cite{Knowles19_Thesis} (figure 3.16). Any differences in the spectra due to [$\alpha$/Fe] are more subtle. Therefore, to try to illustrate any such differences, Figure~\ref{ATK_vs_MaStar_DiffCorr} shows the ratio of enhanced to less-enhanced spectra for averaged MaStar spectra (orange line) of a given type ($\textrm{T}_{\textrm{eff}}$, [Fe/H] and log g) and compares this with the same ratio for the theoretical star spectra (dark blue line). These divisions of spectra represent differential corrections to go from less-enhanced to enhanced spectra. The division of these ratios is shown in the lower panels of each plot in Figure~\ref{ATK_vs_MaStar_DiffCorr}

In Figure~\ref{ATK_vs_MaStar_DiffCorr} some differential features due to [$\alpha$/Fe] changes are qualitatively followed in both the theoretical and observed stars, particularly at short wavelengths where large changes due to abundance pattern variations are seen (e.g. \citealt{Cassisi2004}, their figure 2; \citealt{Sansom2013}, their figure 4; also Figures~\ref{ATK_DiffCorr_m0067} and ~\ref{ATK_DiffCorr_m0923} of this work). However, specific features, such as the region around Mg$_{\textrm{b}}$, show the expected differential behaviour in the theoretical ratio, but this is not well followed by the observed ratio, particularly for the CD stars. A lack of agreement between different SSP models is also found in this broad spectral region, as illustrated in the recent paper by \citealt{Liu2020}, and might be due to uncertainties in MgH molecular band contributions that are particularly important in cool stars. The MaStar spectra that we are comparing our spectral star models to are also likely to suffer from residual continuum differences due to the way that we have had to flatten the spectra in order to be able to compare them with our models. 

Quantitatively, for CG stars, the root-mean-square scatters about unity for the three ratios shown in Figure~\ref{ATK_vs_MaStar_DiffCorr} are: RMS=0.0097, 0.0175, 0.0141 for the ATK ratio, MaStar ratio and (ATK ratio/MaStar ratio) respectively. For CD stars, the corresponding values are: RMS=0.0137, 0.0160, 0.0145 for the ATK ratio, MaStar ratio and (ATK ratio/MaStar ratio) respectively. These values avoided the first and last $200\,${\AA} where continuum fits deviate most. The reductions in RMS values on dividing the two ratios (ATK ratio/MaStar ratio) indicate that the MaStar differential enhancements partially follow the theoretical differential enhancements, but not completely, for both CG and CD stars. Some of the residual mismatches are due to noise in the MaStar data and errors in their abundance estimates. The MaStar CD stars, selected to have the same parameters, show quite a wide range of spectral shapes around the Mg molecular bands and systematic deviations from the theoretical spectrum (Figure~\ref{ATK_vs_MaStar_Spec}, lower panels), suggestive of errors on the [$\alpha$/Fe] measurements of some of those MaStars. This test illustrates the difficulty in testing our theoretically predicted spectral ratios against observations of real stars. The [$\alpha$/Fe] enhancement range available (+0.05 to +0.20 dex) is not much larger than typical errors on [$\alpha$/Fe] enhancements ($\sim\pm$0.1 dex). Large [$\alpha$/Fe] enhancement variations at a given metallicity are not available in the empirical stellar libraries of stars in our Galaxy. 

Given the limitations of empirical star datasets, our match to observed stars seems reasonable, as shown in Figure~\ref{ATK_vs_MaStar_Spec}, for the MaStars selected to be of similar types. In future work improved versions of the MaStar library, with uniform spectral resolution and consistent parameter measurements (rather than heterogeneous ones from the literature, as in \citealt{Yan2019}), may help to improve these comparisons between theoretical and observed star spectra. 

\section{Summary}
\label{sec:Summary}
This work presents new theoretical and semi-empirical stellar spectral libraries, useful for the analysis of stars and stellar populations. 

First, a new high resolution ($\mathrm{R}\sim10^5$) library of theoretical stellar spectra was created to cover a range in stellar parameters including effective temperature, surface gravity, metallicity ($-$2.5$\leq$[M/H]$\leq$+0.5), and covering abundance ratios for $\alpha$-elements ($-$0.25$\leq$[$\alpha$/M]$\leq$+0.75) and carbon ($-$0.25$\leq$[C/M]$\leq$+0.25) (where [M/H]=[Fe/H]). This new library covers parameter ranges of a large proportion of the empirical MILES stars. To minimise the number of models generated, we used an analytical representation of microturbulent velocity as a function of effective temperature and surface gravity based on observational trends found in the literature. These models were generated with consistent abundances of [M/H], [$\alpha$/M] and [C/M] in both their model atmosphere and spectral synthesis components. Existing opacity distribution functions from the APOGEE project were used to create the model atmospheres and the radiative transfer was carried out using ASS$\epsilon$T code in one-dimension, assuming LTE. Kurucz atomic and molecular transitions were included as described in Section~\ref{sec:ModelSpectra}, however, to reduce computation time TiO was excluded for spectra with T$_{\textrm{eff}}>6000${\small K} because it has a negligible effect on stars at these higher temperatures. The resulting theoretical spectra cover a wavelength range from 1680 to $9000\,${\AA} with linear sampling of $0.05\,${\AA} per pixel and are publicly available (see Data Availability Section). 

Comparisons of our new theoretical library with published theoretical spectra from \cite{Allende18} showed good agreement, with small residuals mainly at $\lambda<3000\,${\AA}. Comparisons with PHOENIX models (\citealt{Husser2013}) showed values of Lick indices that generally agreed within typical observational uncertainties on their measurements ($\sim \pm0.1$\,{\AA}) apart from C$_2$4688 and Mg$_\textrm{b}$ indices. We note here that both our models and those of PHOENIX predict a negative change in C$_2$4688 and a positive change in Mg$_\textrm{b}$ for $\alpha$ enhancements (see top panel of Figure ~\ref{fig:ATK_vs_PHOENIX_Lick_Giant}), and therefore both sets of models produce improvements over not considering [$\alpha$/Fe] differential corrections in $\alpha$-enhanced population models. Potential reasons for differences in model predictions for these indices lie in the geometry of underlying atmospheres, as discussed in Section~\ref{ATK_vs_PHOENIX}. Comparing our theoretical spectra directly with MILES empirical spectra highlighted their absolute differences, particularly at bluer wavelengths. Differences are known to be significant between theoretical and empirical star spectra, which is why we have created a library of semi-empirical stellar spectra. Limitations of theoretical models can be explored in future with these new grids.

A differential approach was taken to create a library of semi-empirical stellar spectra covering a range in [$\alpha$/Fe]. Differential corrections were derived from the theoretical grid and applied to empirical star spectra from the MILES library. The resulting grid of semi-empirical (sMILES) model spectra is at the MILES sampling, resolution and wavelength coverage. This library consists of 5 families of 801 semi-empirical star spectra for [$\alpha$/Fe] abundances from $-$0.2 to +0.6 in steps of 0.2 dex. Figure~\ref{sMILES_Library} illustrates the output parameter sampling and coverage, extending the abundance ratios to regions that can be used to model integrated populations from dSphs to giant elliptical galaxies.

 Tests of our new theoretical library against empirical stars from the new MaStar library showed good overall agreement when comparing continuum divided spectra. We tested our predicted differential corrections for [$\alpha$/Fe] variations against ratios of selected cool stars in the MaStar library and found that abundance ratio effects were partially reflected in both, but that cool dwarfs showed a larger range of spectral shapes around the Mg molecular band features. Such tests of our predicted differential corrections are currently limited by the small range in [$\alpha$/Fe] at each metallicity for observed stars in our Galaxy, and by the heterogeneous nature of the MaStar characterisations. Therefore, improved testing awaits better characterisation of MaStar [$\alpha$/Fe] abundances.

Versions of the theoretical and sMILES libraries will be made available on the MILES website for public use. 
\section*{Acknowledgements}
The authors would like to thank the STFC for providing ATK with the studentship for his PhD studies as well as the IAC for providing the support and funds that allowed ATK to visit the institute on two occasions. AES and AV acknowledge travel support from grant AYA2016-77237-C3-1-P from the Spanish Ministry of Economy and Competitiveness (MINECO). AV acknowledges support from grant PID2019-107427GB-C32 from The Spanish Ministry of Science and Innovation. CAP thanks MICINN for grant AYA2017-86389-P.

Funding for the Sloan Digital Sky Survey IV has been provided by the Alfred P. Sloan Foundation, the U.S. Department of Energy Office of Science, and the Participating Institutions. SDSS acknowledges support and resources from the Center for High-Performance Computing at the University of Utah. The SDSS web site is www.sdss.org. SDSS is managed by the Astrophysical Research Consortium for the Participating Institutions of the SDSS Collaboration including the Brazilian Participation Group, the Carnegie Institution for Science, Carnegie Mellon University, the Chilean Participation Group, the French Participation Group, Harvard-Smithsonian Center for Astrophysics, Instituto de Astrofísica de Canarias, The Johns Hopkins University, Kavli Institute for the Physics and Mathematics of the Universe (IPMU) / University of Tokyo, the Korean Participation Group, Lawrence Berkeley National Laboratory, Leibniz Institut für Astrophysik Potsdam (AIP), Max-Planck-Institut für Astronomie (MPIA Heidelberg), Max-Planck-Institut für Astrophysik (MPA Garching), Max-Planck-Institut für Extraterrestrische Physik (MPE), National Astronomical Observatories of China, New Mexico State University, New York University, University of Notre Dame, Observatório Nacional / MCTI, The Ohio State University, Pennsylvania State University, Shanghai Astronomical Observatory, United Kingdom Participation Group, Universidad Nacional Autónoma de México, University of Arizona, University of Colorado Boulder, University of Oxford, University of Portsmouth, University of Utah, University of Virginia, University of Washington, University of Wisconsin, Vanderbilt University, and Yale University.

We also thank the anonymous referee for their comments and suggestions that have greatly improved the clarity and content of this work.

\section*{Data Availability}
The new theoretical stellar spectral library, at fixed spectral sampling, presented in this article are publicly available on the UCLanData repository at {\url{https://uclandata.uclan.ac.uk/178/}}.
The new semi-empirical stellar spectral library will be made publicly available on the MILES website at {\url{http://miles.iac.es/}}. 
%%%%%%%%%%%%%%%%%%%%%%%%%%%%%%%%%%%%%%%%%%%%%%%%%%



\bibliographystyle{mnras}
\bibliography{sMILES_Paper_References}% if your bibtex file is called example.bib


%%%%%%%%%%%%%%%%%%%%%%%%%%%%%%%%%%%%%%%%%%%%%%%%%%

%%%%%%%%%%%%%%%%% APPENDICES %%%%%%%%%%%%%%%%%%%%%

\appendix

\section{Interpolation Choice}
\label{sec:InterpChoice}

\begin{figure*}
\includegraphics[width=120mm,angle=0]{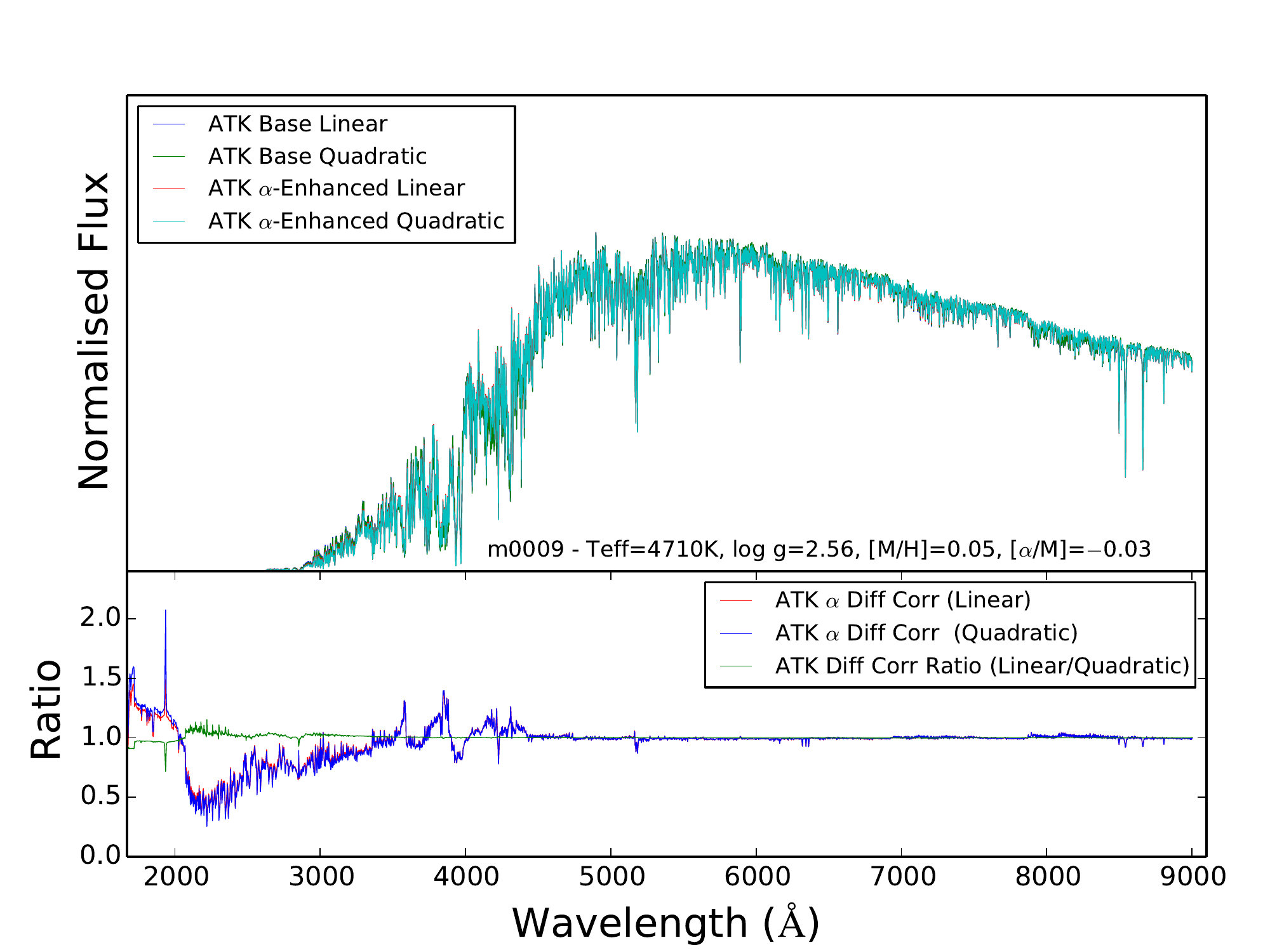}
\caption{Comparison of interpolation methods on differential correction prediction for the full wavelength range of the models. Top plot: Comparison between linear and quadratic interpolation methods on a [$\alpha$/M]=0.25 differential correction for MILES star m0009 (=HD000448). Bottom plot: Predicted [$\alpha$/M]=0.25 differential correction from computed models for a star with parameters close to m0009. All spectra are degraded to MILES resolution of $2.5\,${\AA} and normalised to unity.}
\label{Interpolation_Test_Full}
\end{figure*}

Because we use interpolations within our new theoretical grid to create synthetic MILES stars (see Section~\ref{sec:sMILESstars}), it is important to assess the effect that the interpolation method can have on resulting spectra. We use a differential process, described in Section~\ref{sec:sMILESstars}, to correct empirical MILES stars to account for changes in atmospheric abundances above and below the typical Milky Way abundance pattern in [$\alpha$/Fe]. In this appendix we test the effect of two interpolation methods on the resulting differential predictions,
%, in both the full wavelength range and MILES wavelength range. 

In Figure~\ref{Interpolation_Test_Full} we show the effect of interpolation method on the resulting differential correction through a comparison to computed models across the whole wavelength range of our models. This figure shows the predicted $\alpha$-enhancement differential corrections (ATK $\alpha$ Diff Corr) between a Linear and a Quadratic interpolation method within FER\reflectbox{R}E, for the MILES star m0009. The differential correction is calculated as follows (see Section~\ref{sec:sMILESstars} for a full description of this method). A first interpolation is made within the ATK theoretical grid to create a synthetic MILES star matching the measured $\textrm{T}_{\textrm{eff}}$, log g, [M/H] and [$\alpha$/M] ([Mg/Fe]). A second interpolation is then performed to match the MILES star in $\textrm{T}_{\textrm{eff}}$, log g and [M/H], but with an [$\alpha$/M]=0.25. A ratio of the second and first interpolated spectra gives the predicted differential correction. This correction is a model prediction of how that MILES star will change with an enhancement in atmospheric $\alpha$-element abundances. As shown, there are some differences between the interpolation methods over the full wavelength range, particularly in the UV below $\sim2500\,${\AA}. An RMS scatter around the 1:1 agreement line between linear and quadratic differential corrections for $1677-2500\,${\AA} is 0.193 (see green line). We note however that for the vast majority of the wavelength coverage of the models, the predicted linear and quadratic differential corrections are very similar.

The effect of different interpolation method in the MILES range ($3500-7500\,$\AA) is minimal. An RMS scatter about the 1:1 agreement between linear and quadratic interpolations, for $3500-7500\,${\AA}, is 0.00737. The interpolations also agree very well in the redder wavelengths with RMS scatter of 0.00636 between interpolation methods in the 3500 and $9000\,${\AA} range. We refer interested readers to \cite{Mezaros2013}, who investigate the accuracy of different interpolation methods in both model atmosphere and flux space, for high-resolution optical and infrared stellar spectra.

In conclusion, for the differential application of models to empirical stars in the wavelength range of $3500-9000\,${\AA}, the choice of interpolation method is not important. Both the linear and quadratic interpolation methods in FER\reflectbox{R}E produce similar predictions of the differential correction in this wavelength range. However, there is a non-negligible effect on the predictions at the shorter wavelengths of our models ($<2500\,${\AA}), with significant differences found between linear and quadratic interpolations in this region. For any future applications of the models in the UV, the interpolation method used will have to be considered. Detailed analysis of this wavelength regime is beyond the scope of this current work.

\subsection{11 Stars near Grid Edge}
\label{sec:11probStars}

Interpolations in model grids were performed using the quadratic B\'ezier function within FER\reflectbox{R}E, apart from in 11 low T$_\textrm{eff}$, giant stars. These 11 interpolated spectra showed some negative flux values below $6000\,${\AA}. These theoretical star spectra were:

\begin{itemize}
    \item m0669 - T$_\textrm{eff}$=3640{\small K}, log g=0.70, [Fe/H]=0.00, [Mg/Fe]=0.22
    \item m0704 - T$_\textrm{eff}$=3550{\small K}, log g=0.60, [Fe/H]=0.00, [Mg/Fe]=0.00
    \item m0871 - T$_\textrm{eff}$=3730{\small K}, log g=0.90, [Fe/H]=0.27, [Mg/Fe]=0.27
    \item m0099 - T$_\textrm{eff}$=3560{\small K}, log g=0.60, [Fe/H]=0.00, [Mg/Fe]=0.00
    \item m0234 - T$_\textrm{eff}$=3600{\small K}, log g=0.70, [Fe/H]=-0.30, [Mg/Fe]=0.00
    \item m0614 - T$_\textrm{eff}$=3640{\small K}, log g=0.70, [Fe/H]=-0.10, [Mg/Fe]=0.22
    \item m0481 - T$_\textrm{eff}$=3661{\small K}, log g=1.55, [Fe/H]=0.30, [Mg/Fe]=0.00
    \item m0271 - T$_\textrm{eff}$=3530{\small K}, log g=0.70, [Fe/H]=0.00, [Mg/Fe]=0.00
    \item m0397 - T$_\textrm{eff}$=3700{\small K}, log g=1.22, [Fe/H]=-0.23, [Mg/Fe]=0.22
    \item m0535 - T$_\textrm{eff}$=3600{\small K}, log g=0.70, [Fe/H]=0.00, [Mg/Fe]=0.00
    \item m0053 - T$_\textrm{eff}$=3600{\small K}, log g=1.10, [Fe/H]=0.02, [Mg/Fe]=-0.09
\end{itemize}

\begin{figure}
\centering
 \includegraphics[width=92mm, angle=0]{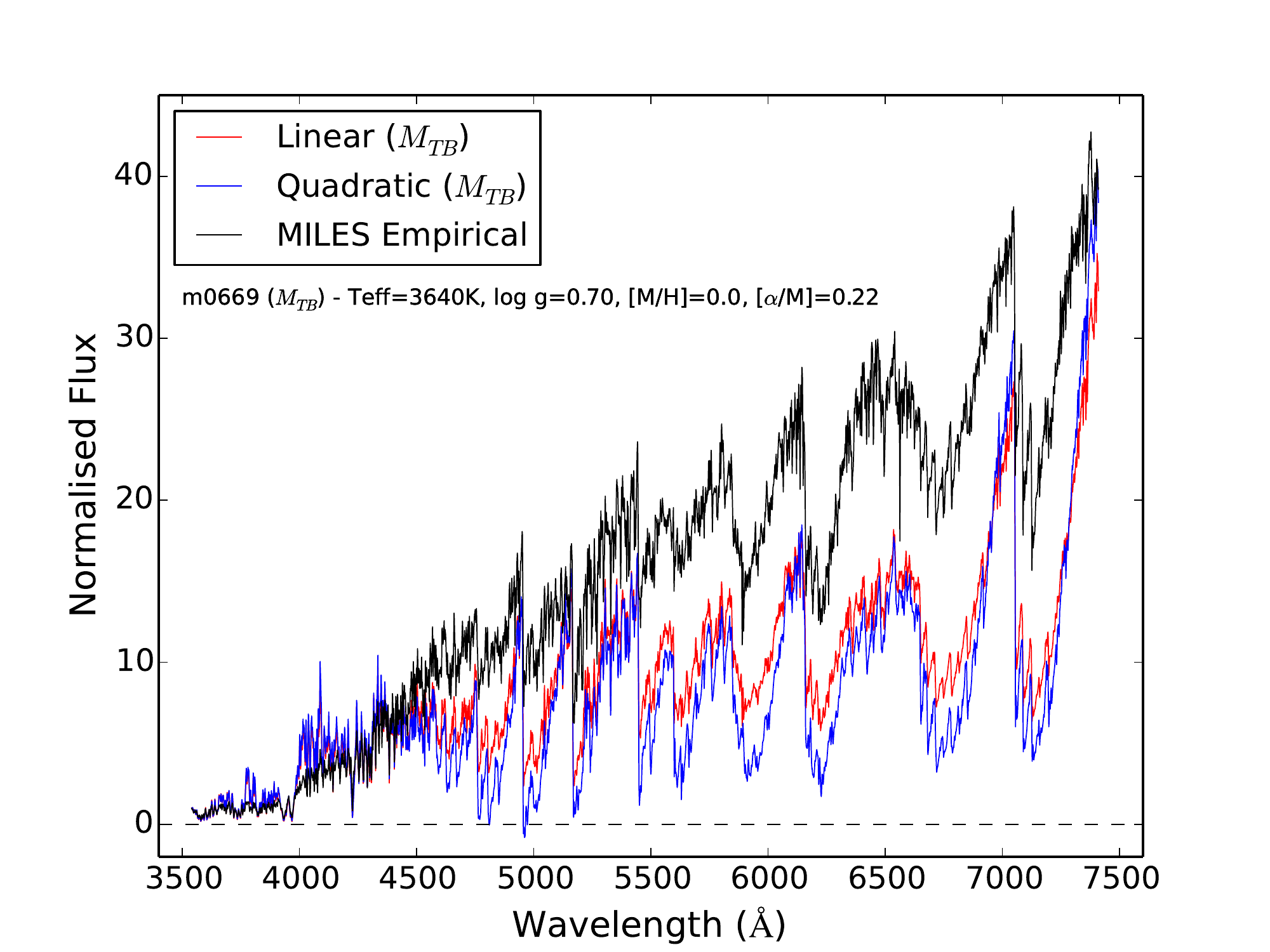}
 \caption[Comparison between spectra produced via linear and quadratic B\'ezier interpolations, near the edges of the theoretical grid] {Comparison between the resulting spectra of a linear and quadratic B\'ezier interpolation, near grid edges, compared to the corresponding empirical spectrum. Spectra are normalised to the flux at $3540.5\,${AA}. The quadratic interpolation produces negative fluxes between $\sim4810-4974\,${\AA}. A linear interpolation fixes this problem and produces a closer match to the corresponding empirical spectrum for m0669 (HD167006, black line).}
 \label{fig:ATK_Linear_vs_Quad_m0669}
\end{figure}
Grid points nearest to these stars were checked for errors, however the problem was found to be with the quadratic interpolation near grid edges. The erroneous spectra were recomputed using a linear interpolation within FER\reflectbox{R}E, with an example of the correction shown in Figure~\ref{fig:ATK_Linear_vs_Quad_m0669}, for the stellar parameters of MILES star m0669 (=HD167006). This example shows negative fluxes for the quadratic interpolation between $\sim4810-4974\,${\AA}, which is improved by using a linear interpolation instead. The spectrum resulting from a linear interpolation better matches the equivalent empirical MILES star spectrum. Similar results were found for the other 10 problematic stars. Other tests were made for several other stars in the library, with the quadratic interpolation found to fit the MILES empirical spectrum better than the linear interpolation in every case.

%\counterwithin{figure}{section}

%\counterwithin{figure}{section}

%\newpage

%%%%%%%%%%%%%%%%%%%%%%%%%%%%%%%%%%%%%%%%%%%%%%%%%%

% Don't change these lines
\bsp	% typesetting comment
\label{lastpage}
\end{document}